\documentclass[reprint,amsmath,amssymb,aps,]{revtex4-1}
\usepackage{amsmath}
\usepackage{graphicx}
\usepackage{tikz}
\usepackage{mathtools}
\usepackage{physics}
\usepackage{bm}
\usepackage[normalem]{ulem}
\usepackage[caption=false]{subfig}
\usepackage{hyperref}
\hypersetup{colorlinks=true,allcolors=blue}
\allowdisplaybreaks

\begin{document}
\title{The $\bm{\beta}$ Fermi-Pasta-Ulam-Tsingou Recurrence Problem}
\author{Salvatore D. Pace, Kevin A. Reiss, and David K. Campbell}
\affiliation{Department of Physics, Boston University\\ Boston, Massachusetts 02215, USA}
\date{\today}

\begin{abstract}
We perform a thorough investigation of the first FPUT recurrence in the $\beta$-FPUT chain for \textit{both} positive and negative $\beta$. We show numerically that the rescaled FPUT recurrence time $T_{r}=t_{r}/(N+1)^{3}$ depends, for large $N$, only on the parameter $S\equiv E\beta(N+1)$. Our numerics also reveal that for small $\left|S\right|$, $T_{r}$ is linear in $S$ with positive slope for both positive and negative
$\beta$. For large $\left|S\right|$, $T_{r}$ is proportional to $\left|S\right|^{-1/2}$ for both positive and negative $\beta$ but with different multiplicative constants. In the continuum limit, the $\beta$-FPUT chain approaches the modified Korteweg-de Vries (mKdV) equation, which we investigate numerically to better understand the FPUT recurrences on the lattice. In the continuum, the recurrence time closely follows the $|S|^{-1/2}$ scaling and can be interpreted in terms of solitons, as in the case of the KdV equation for the $\alpha$ chain. The difference in the multiplicative factors between positive and negative $\beta$ arises from soliton-kink interactions which exist only in the negative $\beta$ case. 
We complement our numerical results with analytical considerations in the nearly linear regime (small $\left|S\right|$) and in the highly nonlinear regime (large $\left|S\right|$). For the former, we extend previous results using a shifted-frequency perturbation theory and find a closed form for $T_{r}$ which depends only on $S$. In the latter regime, we show that $T_{r}\propto\left| S\right|^{-1/2}$ is predicted by the soliton theory in the continuum limit. We end by discussing the striking differences in the amount of energy mixing as well as the existence of the FPUT recurrences between positive and negative $\beta$ and offer some remarks on the thermodynamic limit.
\end{abstract}

\maketitle

\section{Introduction}
Fermi, Pasta, Ulam, and Tsingou (FPUT) numerically investigated the dynamics of a homogeneous anharmonic chain initialized far from equilibrium to study the effect of nonlinear interactions on thermal conductivity and ergodicity in solids \cite{Fermi1955}. To their surprise, their computational results showed that within observed time scales, the system did not obtain energy equipartition among normal modes. Instead, when the first normal mode was initially excited, they observed energy flow among only the lowest normal modes with periodic near-recurrences to the initial state. These \textit{FPUT recurrences} and the questions they raise about how systems approach equilibrium have puzzled and challenged researchers for more than sixty years \cite{Ford1992,Weissert,Berman2005, Gallavotti}.

Since their discovery, three seemingly different pictures have been developed to explain the cause of the recurrences in what is now called the FPUT chain. First revealed by a shifted-frequency perturbation theory, the FPUT recurrences occur because of near resonances between perturbatively defined nonlinear frequencies  \cite{Ford1961,Sholl1991}. Next, in an interpreted continuum limit, the equations of motion become a partial differential equation with solitary wave solutions called \textit{solitons}. In this limit, FPUT's initial conditions break up into multiple solitons moving at different  velocities, and the FPUT recurrence occurs when the solitons overlap in the same configuration they started in \cite{Zabusky1965}. Finally, the third explanation comes from time-periodic localized structures in normal mode space called $q$-breathers \cite{Flach2005}, or more generally $q$-tori \cite{Christodoulidi2010}, which are exact solutions of the FPUT chain's equations of motion. The FPUT recurrences occur because the trajectories of long-wavelength initial excitations are perturbations of the $q$-tori orbits.

Much of the previous work---in particular, that dealing with the soliton approach---has focused on the $\alpha$-FPUT chain, in which the nonlinear term in the Hamiltonian is proportional to the cube of the difference in the positions of neighboring masses. In the present work, we investigate the FPUT recurrence and its scaling in the $\beta$-FPUT chain, in which the nonlinear term in the Hamiltonian is proportional to the fourth power of the difference in the positions of neighboring masses. Indeed, while there have been numerous studies on the FPUT recurrences in the $\alpha$-FPUT chain \cite{Zabusky1963, Zabusky1969, Toda1969, Lin1997}, to our knowledge, there has yet to be a complete study on them in the $\beta$-FPUT chain. In later sections of this article, we offer detailed reasons why this situation has occurred but in brief, it has to do with instabilities known to exist both on the lattice and in continuum limit \cite{Driscoll1976}, which make numerical studies of the $\beta$-FPUT recurrences challenging \cite{Drago1987, Pace2019}, as well as peculiarities in the method of solution for the continuum model. We will clarify these somewhat cryptic comments later.

In the majority of previous studies of the $\beta$-FPUT chain, the quartic interactions were attractive ($\beta>0$) rather than repulsive ($\beta<0$). This is likely because negative $\beta$ causes saddle points in the potential and therefore, \textit{large enough} initial energies will lead to trajectories blowing up. The positive $\beta$-FPUT's potential is instead completely bounded from below. However, it is still interesting to study unbounded potentials for energies below the saddle points. Indeed, in the $\alpha$-FPUT chain, such saddle points always exist because the nonlinear interaction is cubic. Moreover, as noted by previous studies \cite{DeLuca1995}, for energies below any blow-up, the dynamics for $\beta<0$ are not a trivial extension of the $\beta>0$ results and should be independently considered. For example, the solitons which form in the continuum limit for $\beta>0$ are exponentially unstable, whereas for  $\beta<0$, they are (perhaps surprisingly) stable \cite{Driscoll1976}. Furthermore, the dynamics of a one-dimensional Bose gas in the quantum rotor regime maps onto the $\beta$-FPUT chain with $\beta<0$ \cite{Danshita2014}. Hence, a primary focus of our study will be the comparison of the FPUT recurrences between the conventional $\beta>0$ and atypical $\beta<0$.

We note that along with their central role in FPUT chains, FPUT recurrences have also been found and studied in numerous other theoretical models and also observed experimentally in various physical systems. Examples of the former include electron-phonon interactions \cite{Kopidakis1994}, interacting Bose-Einstein condensates \cite{Danshita2014, Osanov2015}, ion waves \cite{Abe1981}, and anti-de Sitter spacetime \cite{Balasubramanian2014,Biasi2019}. The latter involves experimental demonstrations with electrical LC networks \cite{Hirota1970}, plasmas \cite{Ikezi1973}, deep-water waves \cite{Lake1977}, silica optical fibers \cite{VanSimaeys2001}, magnetic film strips \cite{Scott2003,Wu2007}, and optics in a photorefractive crystal \cite{Pierangeli2018}.

The outline of the paper is as follows. In section \ref{chain_and_cont}, we introduce the $\beta$-FPUT chain and show how its continuum limit becomes the well-known modified Korteweg-de Vries (mKdV) equation. Then, in section \ref{sec:numerics_beta_rec_scale} we report and discuss our numerical determination of the scaling in time of  the FPUT recurrences, importantly showing that a rescaled FPUT recurrence time depends, for large $N$,  only on the parameter $S\equiv E\beta(N+1)$. 
In section \ref{sec:nearly_linear}, we extend the results from reference \cite{Sholl1991} and present analytic results for the FPUT recurrence time in the nearly linear regime as a function of $S$ only. 
In section \ref{sec:mkdv}, we investigate numerically the recurrence of the initial state in the mKdV equation and examine the role played by solitons. Next, in section \ref{NLR}, we explore the highly nonlinear regime and find an analytical expression for the FPUT recurrence time using the soliton velocities in the continuum. In section \ref{sec:rec_existence}, we look at the dependence of the formation of the first FPUT recurrence on lattice parameters. Finally, in section \ref{sec:conclusion}, we offer concluding remarks and highlight some open questions.

\section{The $\bm{\beta}$-FPUT Chain and its Continuum Limit}\label{chain_and_cont}

The $\beta$-FPUT chain is described by the classical Hamiltonian
\begin{equation}\label{beta_hamiltonian}
H = \sum_{n=0}^{N}\frac{p_{n}^{2}}{2m}+\frac{k_{2}}{2}(q_{n+1}-q_{n})^{2}+\frac{k_{4}}{4}(q_{n+1}-q_{n})^{4},
\end{equation}
where $q_{n}$ and $p_{n}$ are, respectively, the displacement from rest and canonical momentum of site $n$, $N$ is the number of active masses, $m$ is mass, and $k_{2}$ and $k_{4}$ are weights of the quadratic and quartic interactions, respectively. Using the canonical transformation
\begin{equation}\label{canonical_transformation}
\begin{bmatrix}
q_{n}\\
p_{n}
\end{bmatrix}
=
\sqrt{\frac{2}{N+1}} \sum_{k=1}^{N}
\begin{bmatrix}
Q_{k}\\
P_{k}
\end{bmatrix}
\sin \left(\frac{n k \pi}{N+1}\right),
\end{equation}
we diagonalize the quadratic term of the Hamiltonian and transform to normal mode coordinates $\left(Q_{k},P_{k}\right)$. With this, equation (\ref{beta_hamiltonian}) becomes
\begin{equation}\label{beta_NM_hamiltonian}
H = \sum_{k=1}^{N}\frac{P_{k}^{2}}{2m}+\frac{k_{2}\omega_{k}^{2}Q_{k}^{2}}{2}+\frac{k_{4}}{4}\sum_{i,j,l=1}^{N}C_{kijl}Q_{k}Q_{i}Q_{j}Q_{l},
\end{equation}
where the normal mode frequencies are
\begin{equation}\label{NM_frequency}
\omega_{k}=2\sin\left(\frac{k\pi}{2(N+1)}\right)
\end{equation}
and the coupling constants are \cite{Sholl1990}
\begin{equation}\label{coupling_constant}
C_{kijl} = \frac{\omega_{k}\omega_{i}\omega_{j}\omega_{l}}{2(N+1)}\sum_{ \pm}\left[\delta_{k, \pm j \pm l \pm m}-\delta_{k \pm j \pm l \pm m, \pm 2(N+1)}\right].
\end{equation}
The sum $\sum_{\pm}$ in equation (\ref{coupling_constant}) is over all possible combinations of plus and minus signs in the $\pm$s of the Kronecker delta functions, $\delta_{i,j}$.

From Hamilton's equations $\left(\dot{q} = \partial_{p}H, \dot{p} = -\partial_{q}H\right)$, the real space equations of motion are
\begin{equation}\label{beom}
\ddot{q}_{n}\hspace{-2pt}=q_{n+1}+q_{n-1}-2q_{n}+b\left|\beta\right|\hspace{-1pt}\left[ (q_{n+1}\hspace{-1pt}-\hspace{-1pt}q_{n})^{3}\hspace{-1pt}-\hspace{-1pt}(q_{n}\hspace{-1pt}-\hspace{-1pt}q_{n-1})^{3} \right],
\end{equation}
where we have defined $\beta\equiv k_{4}/k_{2}$ and rescaled time by the multiplicative constant $\sqrt{k_{2}/m}$. Furthermore, because we will be considering both $\beta>0$ and $\beta<0$, we have taken the absolute value of $\beta$ and have introduced $b\equiv\text{sgn}(\beta)$. In the continuum limit, in which $N$ goes to infinity while the lattice spacing, $a$, goes to zero such that the length of our chain, $L=a(N+1)$, is fixed, the displacement variables $\{q_{n}(t)\}$ become a field $q(x,t)$, with 
\begin{equation}
q_{n}(t)\equiv q(na,t).
\end{equation}
To take the continuum limit, we Taylor expand $q_{n\pm 1}(t)$ to fourth order in $a$,
\begin{equation}
q_{n\pm 1}(t) = q\pm a q_{x}+\frac{a^{2}}{2}q_{xx}\pm\frac{a^{3}}{6}q_{xxx}+\frac{a^{4}}{24}q_{xxxx},
\end{equation}
where we employ the notation $q(x,t)\equiv q$ and use subscripts to denote partial differentiation. Plugging this in the equations of motion, dropping fifth order terms in $a$, and then simplifying we find
\begin{equation} \label{beta_cont}
\ddot{q}=a^{2}\left( q_{xx}+b\varepsilon(q_{x})^{2}q_{xx}+\zeta \varepsilon q_{xxxx}\right),
\end{equation}
where we have defined the variables $\varepsilon = 3\left|\beta\right| a^{2}$ and $\zeta=1/(36\left|\beta\right|)$. 
Let us consider the asymptotic solution of a right-going wave, $q(x,t) \sim F(\xi,\tau)$ where $\xi=x-at$ and $\tau = \varepsilon at/2$ \cite{Albowitz}. Plugging this into equation (\ref{beta_cont}) gives
\begin{equation}
F_{\xi, \tau}+b (F_{\xi})^{2}F_{\xi\xi} +\zeta F_{\xi\xi\xi\xi} =0.
\end{equation}
Introducing a new field $\phi = F_{\xi}$, we arrive at the 
standard form for the mKdV equation
\begin{equation}
\label{mkdvfinal}
\phi_{\tau}+b\phi_{\xi}\phi^{2}+\zeta\phi_{\xi\xi\xi}=0.
\end{equation}
In the literature, the mKdV equation is called the ``focusing'' (``defocusing'') mKdV equation when $b=1$ ($b=-1$).

The initial conditions on the lattice are $q_{n}(0)=A\sin\left(n\pi/(N+1)\right)$, and we will use fixed boundary conditions, $q_{0}=q_{N+1}=0$. Because the equations of motion of the $\beta$-FPUT chain preserve odd symmetry about the chain's center, the dynamics of our chain will be equivalent to a chain with $2N+2$ active masses and the same initial conditions but under the periodic boundary conditions $q_{0}=q_{2N+2}$ \cite{Zabusky1963}. In the continuum limit, we will take advantage of this and consider $\xi\in\left[0,2L\right]$ with $\phi(0,\tau) = \phi(2L,\tau)$. The initial conditions on the lattice will become $\phi(\xi,0) \sim q_{x}(x,0)/2 = A\kappa\cos(\kappa \xi) /2$, where $\kappa = \pi/L$. We have divided by two to take into account that the mKdV equation only describes the right-going waves.

\section{Numerical Scaling of FPUT Recurrence Time}\label{sec:numerics_beta_rec_scale}

In this section, we report our numerical results for the dependence of the FPUT recurrence time on the parameters of the $\beta$-FPUT chain ($N$, $\beta$, and the energy $E$). Our numerics were performed using the symplectic SABA$_{2}$C integrator \cite{Laskar2001}. This numerical scheme produces an error of $\mathcal{O}\left(dt^{4}\right)$, where $dt$ is the numerical time step size, which we typically have set to $dt=0.1$. For more details regarding the relative energy error and the explicit form of the numerical scheme for FPUT-chains, see the appendix of reference \cite{Pace2019}. 

\begin{figure}[t!]
		\centering
		\includegraphics[width=.48\textwidth]{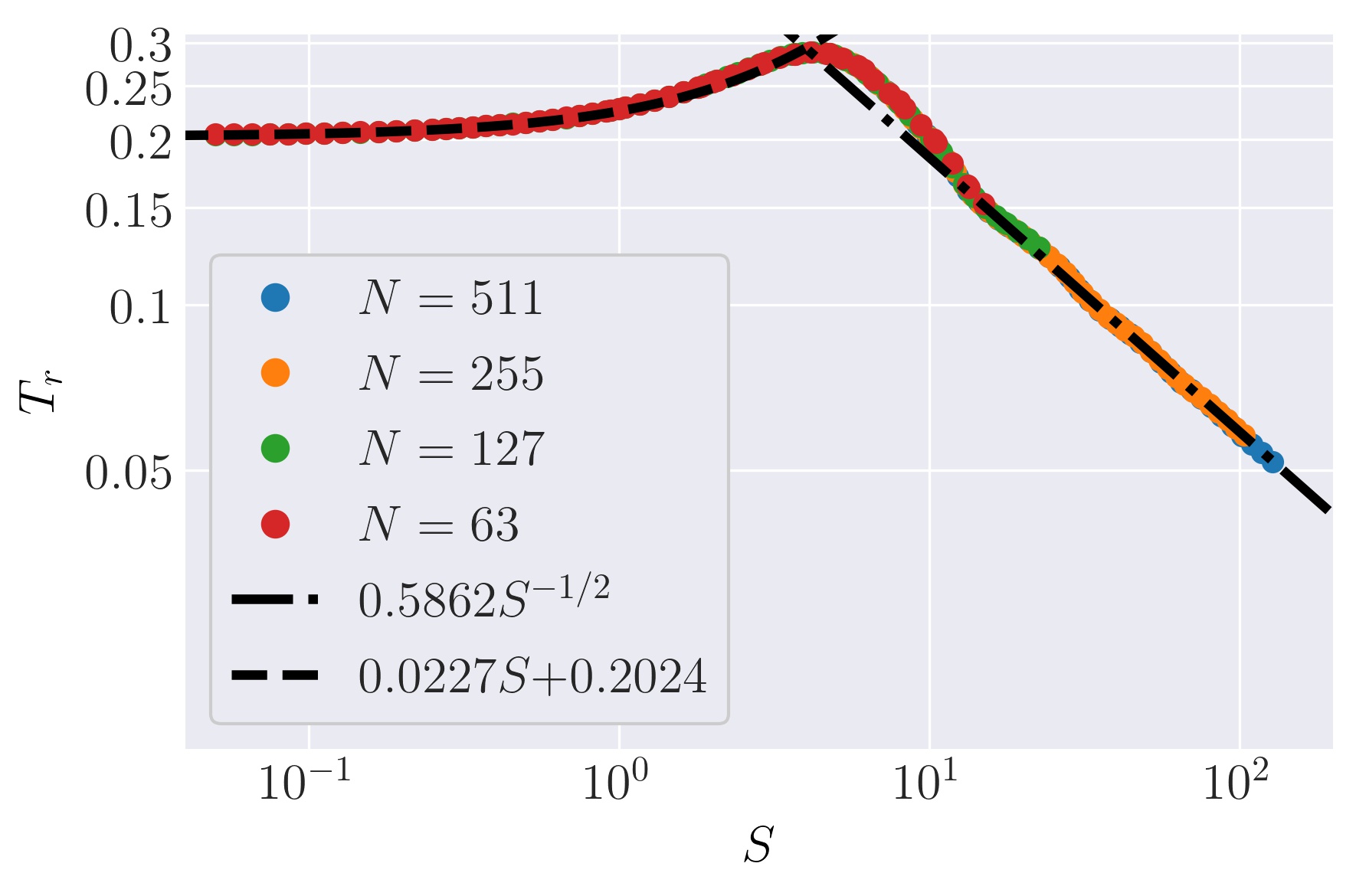}
		\caption{(color online) Numerical data for the rescaled FPUT recurrence time $T_{r}\equiv t_{r}/(N+1)^{3}$ as a function of $S\equiv E\beta(N+1)$ on a log-log plot for the $\beta$-FPUT chain with $\beta>0$. For the largest system size, the data extend over the full range of $S$ shown in the plot. For smaller sizes, the data stop at certain values of $S$ for reasons discussed in the text. The two dashed lines are numerically generated fits.} 
		\label{pos_beta_rec_scale}
\end{figure}

In the $\alpha$-FPUT chain, for large $N$, the rescaled recurrence time $T_{r}\equiv t_{r}/(N+1)^{3}$ depends only on the parameter $R\equiv E\alpha^{2}(N+1)^{3}$ \cite{Lin1997, Zabusky1969, Toda1975}. Rescaling the recurrence time by $(N+1)^{3}$ comes from the perturbative result that in the large $N$ limit, as the chain approaches the harmonic limit the FPUT recurrence time scales like $(N+1)^{3}$ \cite{Ford1961}. While considering a subsystem restricted to only four consecutive normal modes of the $\beta$-FPUT chain with $\beta>0$, De Luca \textit{et al.} found that a limiting form of the resulting Hamiltonian depends only on the parameter $6\beta E(N+1)/\pi^{2}$ \cite{DeLuca1995}. Thus, it is reasonable to investigate the rescaled FPUT recurrence times in the $\beta$-FPUT chain as a function of the parameter $S\equiv E\beta (N+1)$. This parameter importantly takes into account that we expect there should be a difference in the dynamics based on the sign of $\beta$.

\begin{figure}[t!]
		\centering
		\includegraphics[width=.48\textwidth]{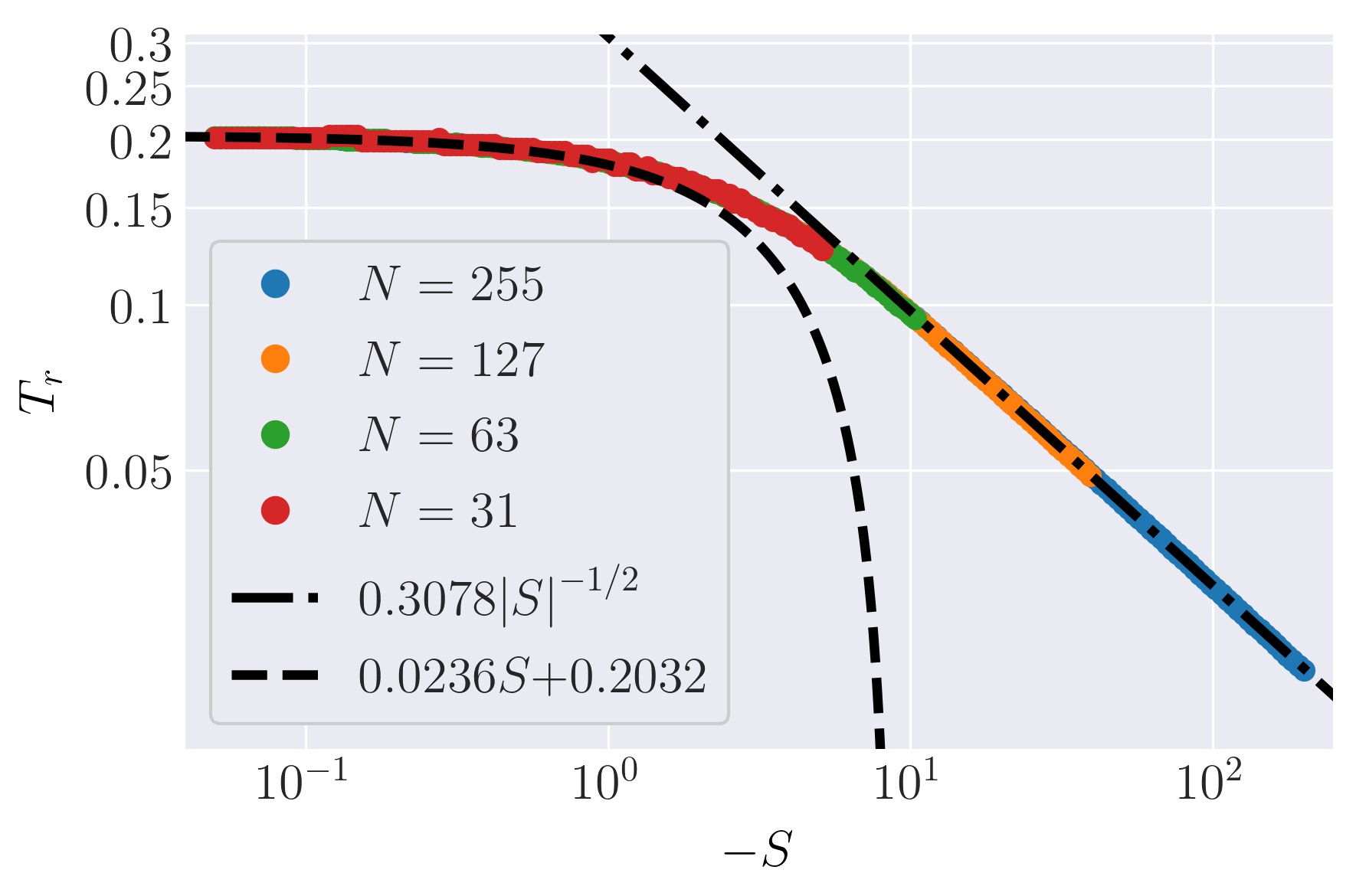}
		\caption{(color online) Numerical data of the rescaled FPUT recurrence time $T_{r}\equiv t_{r}/(N+1)^{3}$ as a function of $S\equiv E\beta(N+1)$ on a log-log plot for the $\beta$-FPUT chain with $\beta<0$. For the largest system size, the data extend over the full range of $-S$ shown in the plot. For smaller sizes, the data stop at certain values of $-S$ for reasons discussed in the text. The two dashed lines are numerically generated fits.}  
		\label{neg_beta_rec_scale}
\end{figure}

Let us first consider the more conventional $\beta>0$ case. Figure \ref{pos_beta_rec_scale} shows our numerical results on a log-log plot for the rescaled recurrence time, $T_{r}$ as a function of $S$. We start in the regime of small $\left|S\right|$ and fit our data with a linear function. We find that for $0<S\lesssim 1$, the data goes like
\begin{equation}\label{small_pos_beta}
T_{r} = 0.2024+0.0227S.
\end{equation}
For large $\left|S\right|$, as is clear from the plot, the rescaled FPUT recurrence time exhibits power-law scaling. Fitting the numerical data gives the expression
\begin{equation} \label{eqn:pos_beta_scale}
T_{r} = \frac{0.5862}{\sqrt{S}},
\end{equation}
which provides an accurate fit for $S\gtrsim 30$.

Moving on to $\beta<0$, Figure \ref{neg_beta_rec_scale} shows the numerical results of the rescaled FPUT recurrence time. For small $\left|S\right|$, we again fit our data with a linear function and find that for $-0.5\lesssim S<0$,
\begin{equation}\label{small_neg_beta}
T_{r} = 0.2032+0.0236S.
\end{equation}
For large $\left|S\right|$, we find that once again the rescaled recurrence time shows power law scaling. Numerically fitting this region gives that for $S\lesssim -12$, 
\begin{equation} \label{eqn:neg_beta_scale}
T_{r} = \frac{0.3078}{\sqrt{\left|S\right|}}.
\end{equation}

So, for both positive and negative $\beta$, the rescaled FPUT recurrence time at large $|S|$ is proportional to $|S|^{-1/2}$. For small $\left|S\right|$, the rescaled FPUT recurrence time scales linearly, smoothly transitioning between negative and positive $\beta$. Indeed, comparing equations (\ref{small_pos_beta}) and (\ref{small_neg_beta}) shows that the numerical fits in this region are essentially independent of the sign of $\beta$. Furthermore, it is interesting to observe that 
the maximum value of $T_{r}$ is not at zero but instead at $S\sim 4.21$.

Note that in both Figures \ref{pos_beta_rec_scale} and \ref{neg_beta_rec_scale}, the numerical results for a given $N$ are not plotted above some value of $\left|S\right|$. The reasons for this differ between the two cases. When $\beta>0$, this is because the FPUT recurrences simply stop forming; we will discuss this in greater detail in section \ref{sec:rec_existence}. When  $\beta<0$, for a given system size $N$,
for large enough $\left|S\right|$ the rescaled FPUT recurrence times develop an $N$ dependence and hence start to disagree with Figure \ref{neg_beta_rec_scale}. 
However, for large enough system size $N$, the dynamics on the lattice are well modeled by the dynamics in the continuum and indeed depend only on the parameter $S$.  This will be evident in sections \ref{sec:nearly_linear} and \ref{NLR} where our analytical results rely asymptotically on only $S$ when $N$ is large. In the $\alpha$-FPUT chain, the threshold between ``small'' and ``large'' system sizes, and the energy at which the chain can start blowing up due to the unbounded potential are correlated \cite{C.Y.Lin1999}. As we show in the Appendix \ref{sec:beta_large_vs_small}, the small versus large threshold for the $\beta$-FPUT chain with $\beta<0$ follows the same correlation and that the system can be considered large when $N> 4\left|S\right|-3$.

Therefore, we see that as in the case of $R$ for the $\alpha$-FPUT chain, for large $N$, the parameter $S$ fully describes the rescaled FPUT recurrence time in the $\beta$-FPUT chain for both positive and negative $\beta$. Interestingly, while we have shown in a previous study that $R$ describes the singularities in the periods of the higher-order recurrences (e.g., super recurrences, super-super recurrences, etc.) in the $\alpha$-FPUT chain, we have not found that $S$ describes such singularities in the $\beta$-FPUT chain \cite{Pace2019}.

\section{Nearly Linear Regime Analytics}\label{sec:nearly_linear}

When $\left|S\right|$ is small, the system is in a nearly linear regime, and its dynamics can be understood by adding perturbative corrections to the harmonic chain's dynamics and acoustic phonon frequencies. This so-called ``shifted-frequency perturbation theory" has been used to study the dynamics of both the $\alpha$ and $\beta$ FPUT-chain for first normal mode initial excitations \cite{Ford1961,Sholl1991}. In this section, we look at the expression \begin{equation}\label{SH_recc}
t_{r} = \frac{2\pi}{3\Omega_{1}-\Omega_{3}},
\end{equation}
found by Sholl and Henry for the FPUT recurrence time in the $\beta$-FPUT chain \cite{Sholl1991}. The $\Omega_{k}$s are nonlinear frequencies which have been calculated to second order in $\beta$. For more information about the perturbation scheme and how the nonlinear frequencies are constructed, we refer the reader to Appendix \ref{Sec:PT}.

As shown numerically in section \ref{sec:numerics_beta_rec_scale}, the rescaled FPUT recurrence time for large $N$ is a function of only $S\equiv E\beta(N+1)$. Therefore, at large $N$, the perturbative results should become only a function of $S$ as well. Let us first rewrite the perturbative parameter, $\beta$, as a function of $S$ and $N$. Sholl and Henry considered the initial conditions $Q_{k}(0) = \delta_{k,1}$, which corresponds to real space initial conditions $q_{n} = \sqrt{2/(N+1)}\sin(n\pi/(N+1))$. Using equation (\ref{beta_initial_S}) from appendix \ref{amps}, this corresponds to initial conditions with
\begin{equation}\label{SH_S}
S = \frac{\omega_{1}^{2}\beta(N+1)}{2}\left(1+\frac{3\beta\omega_{1}^{2}}{4(N+1)}\right).
\end{equation}

Therefore, by solving for $\beta$ in the above equation, we can replace the $\beta$s in the definition of the nonlinear frequencies and find the FPUT recurrence time as a function of only $S$ and $N$. After some involved calculations with Mathematica\texttrademark, we find that the denominator of equation (\ref{SH_recc}) can be written as
\begin{equation}
\begin{aligned}
3\Omega_{1}-\Omega_{3} &= \frac{1}{(N+1)^{3}}\bigg(\frac{46 \pi ^3}{3}+\frac{15 \pi  S}{16}+\frac{240 \pi ^5 }{3 S-16 \pi ^2}\\
&\hspace{20pt}+\frac{16 \pi ^5 }{24 \pi ^2-27 S}\bigg)+\mathcal{O}\left(\frac{1}{(N+1)^{5}}\right).
\end{aligned}
\end{equation}
Dropping the 5th order terms, inserting this into equation (\ref{SH_recc}), and then rearranging terms, we find that the rescaled FPUT recurrence time for small $S$ is given by
\begin{equation}\label{pert_rec_time}
T_{r} = \frac{864 S^{2}-5376 \pi^{2} S+4096 \pi^{4}}{405S^{3}+4104 \pi^{2} S^{2}-4992 \pi^{4} S+2048 \pi^{6}}.
\end{equation}

 \begin{figure}[t!]
		\centering
		\includegraphics[width=.48\textwidth]{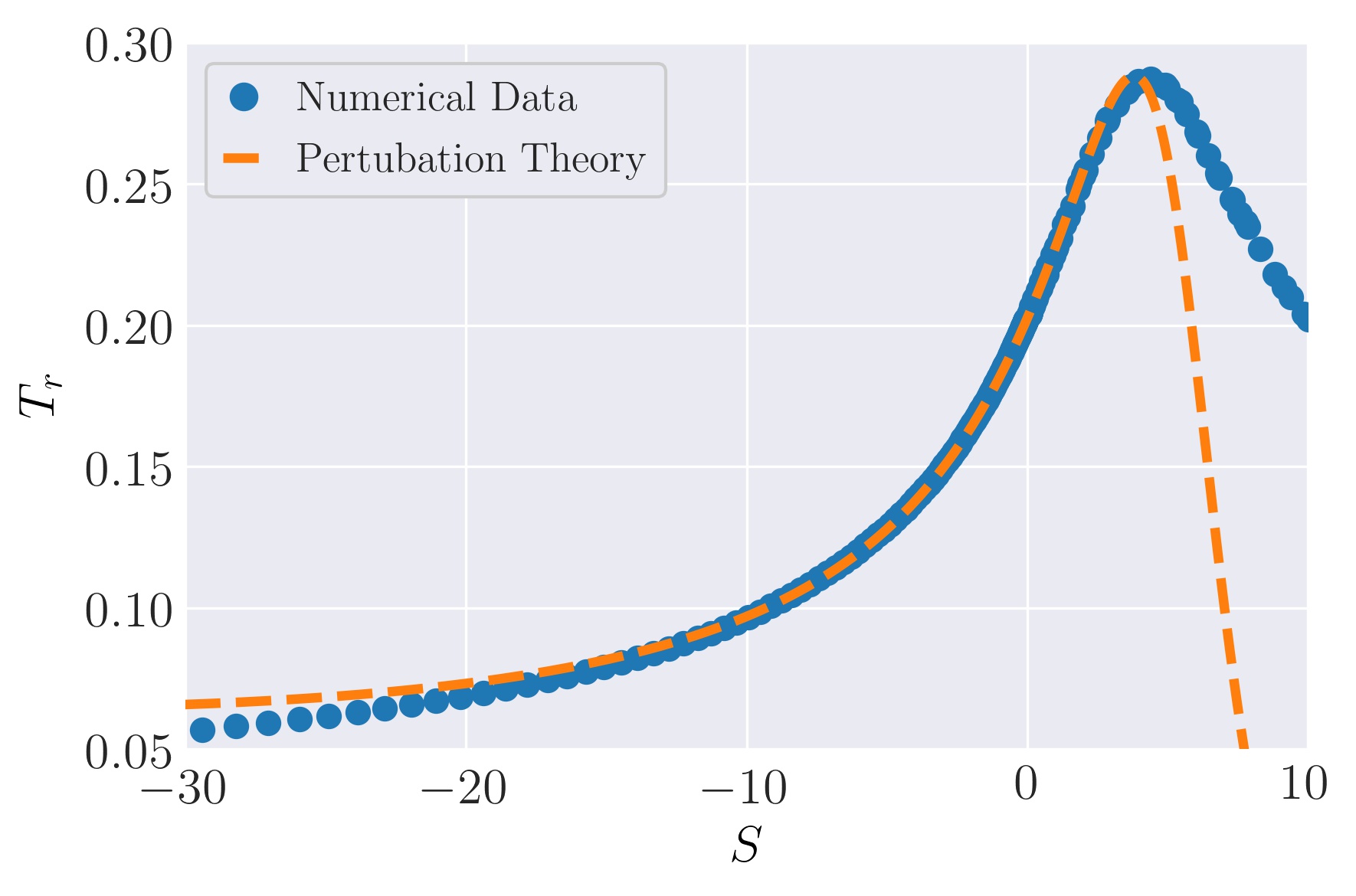}
		\caption{The numerical data for the rescaled FPUT recurrence time plotted along with the results of equation (\ref{pert_rec_time}) which was obtained analytically from the ``shifted-frequency perturbation theory" \cite{Sholl1991}.} 
		\label{SH_analytic_beta_rec_scale}
\end{figure}

This expression for $T_{r} $ is shown in Figure \ref{SH_analytic_beta_rec_scale}, along with the numerically determined rescaled FPUT recurrence time for large systems. It can be seen that equation (\ref{pert_rec_time}) is an accurate expression for the rescaled recurrence time in the domain $-12\lesssim S\lesssim 4$. Writing equation (\ref{pert_rec_time}) to first order in $S$ yields
\begin{equation}
\begin{aligned}
T_{r} &= \frac{2}{\pi^{2}} + \frac{9}{4\pi^{4}}S+\mathcal{O}(S^{2})\\
&\sim 0.2026+0.0231S .
\end{aligned}
\end{equation}
Comparing this to the numerical linear fits for small $\left|S\right|$ given in equations (\ref{small_pos_beta}) and (\ref{small_neg_beta}) shows excellent agreement.

\begin{figure*}
	\subfloat[]{\includegraphics[width=.48\textwidth]{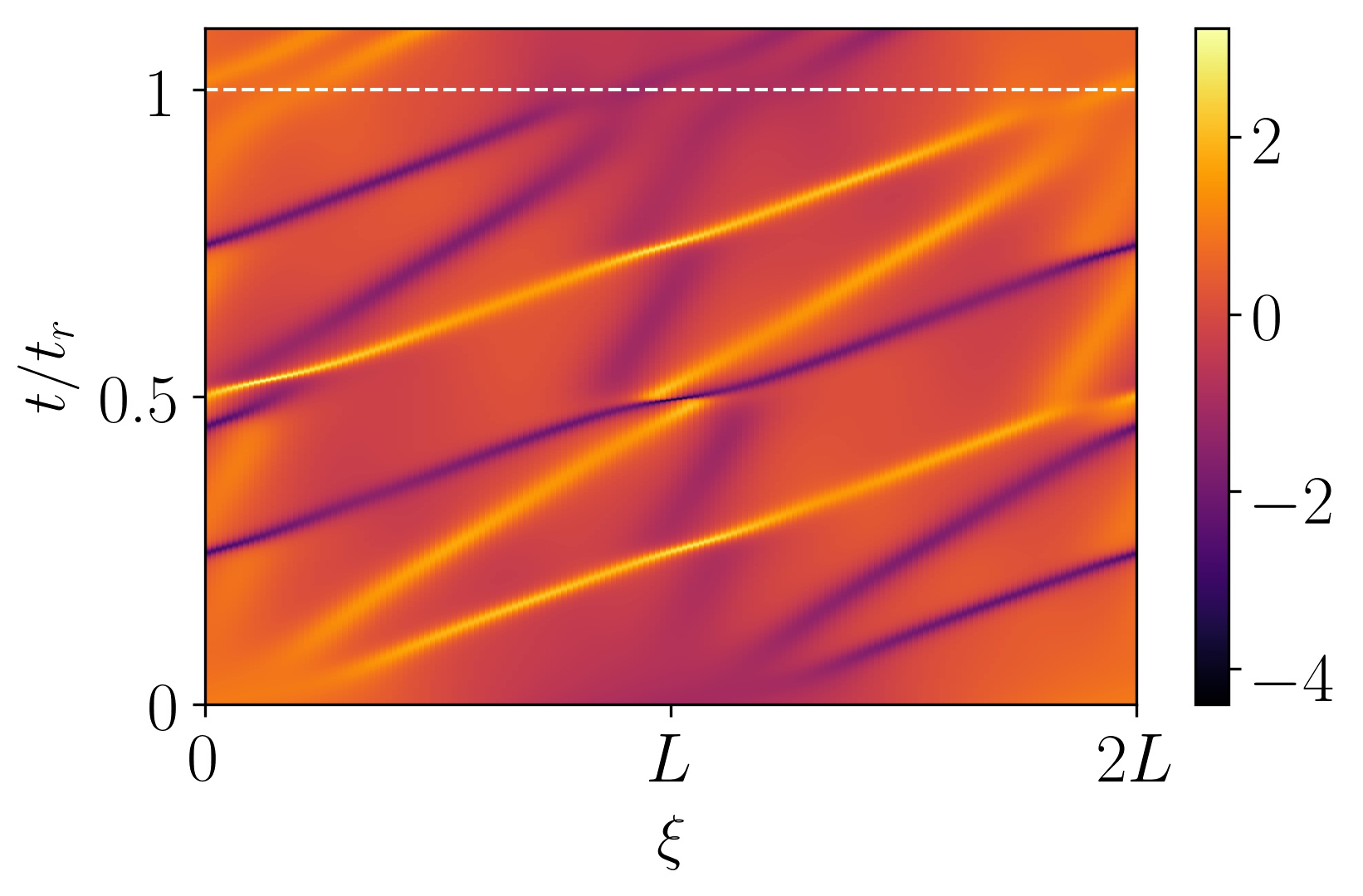}}
	\subfloat[]{\includegraphics[width=.48\textwidth]{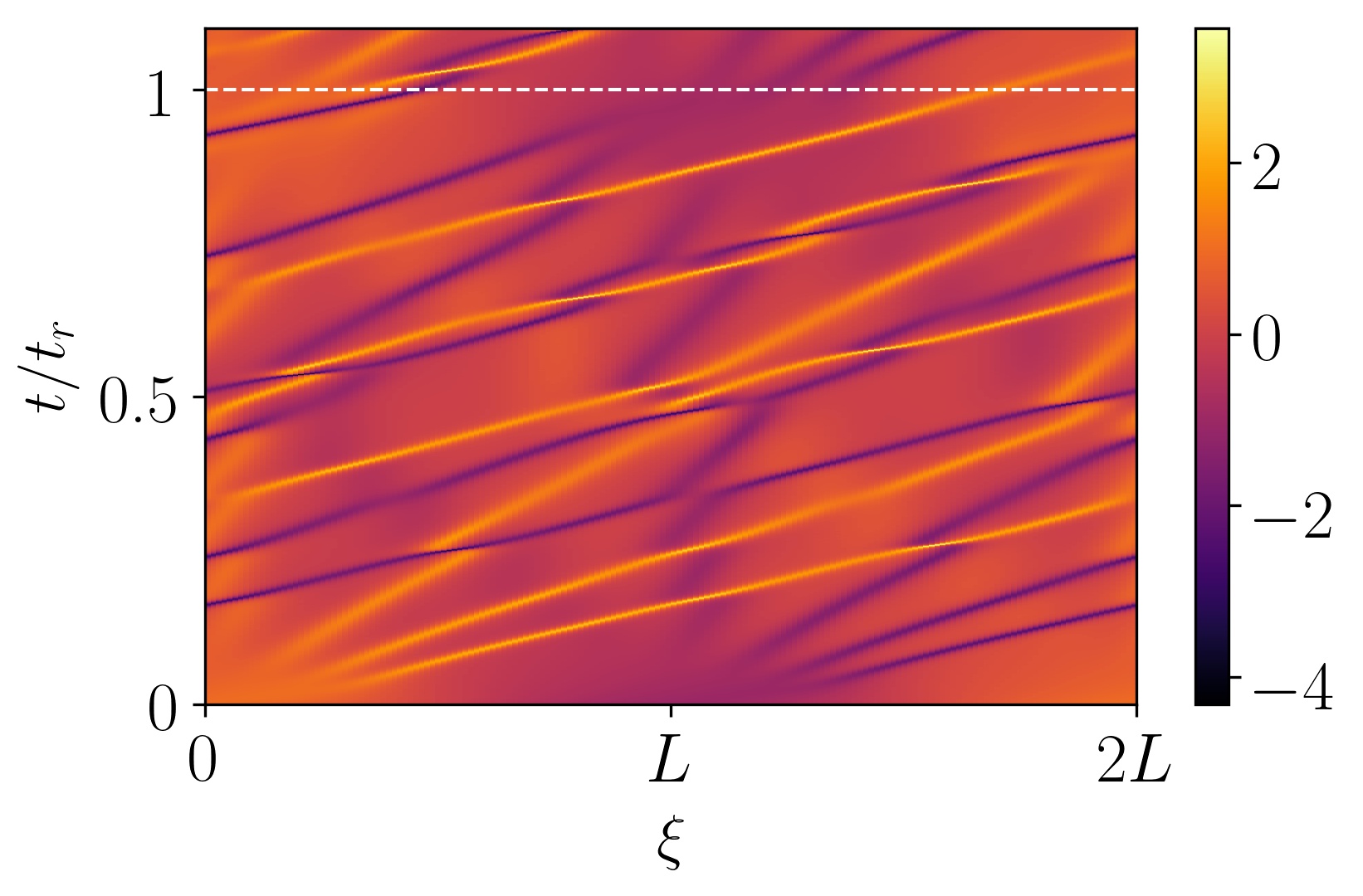}}\hfill
	\subfloat[]{\includegraphics[width=.48\textwidth]{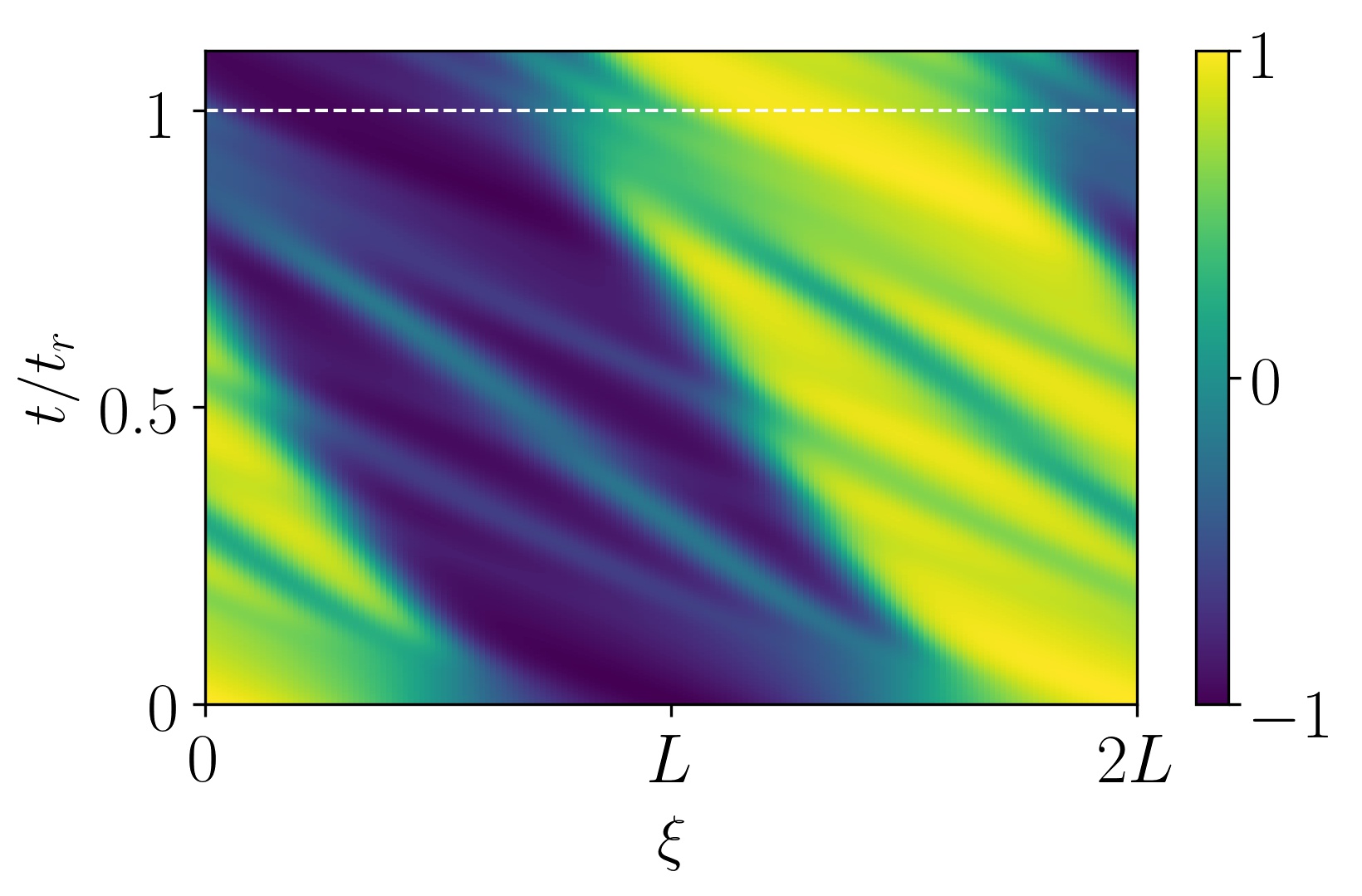}}
	\subfloat[]{\includegraphics[width=.48\textwidth]{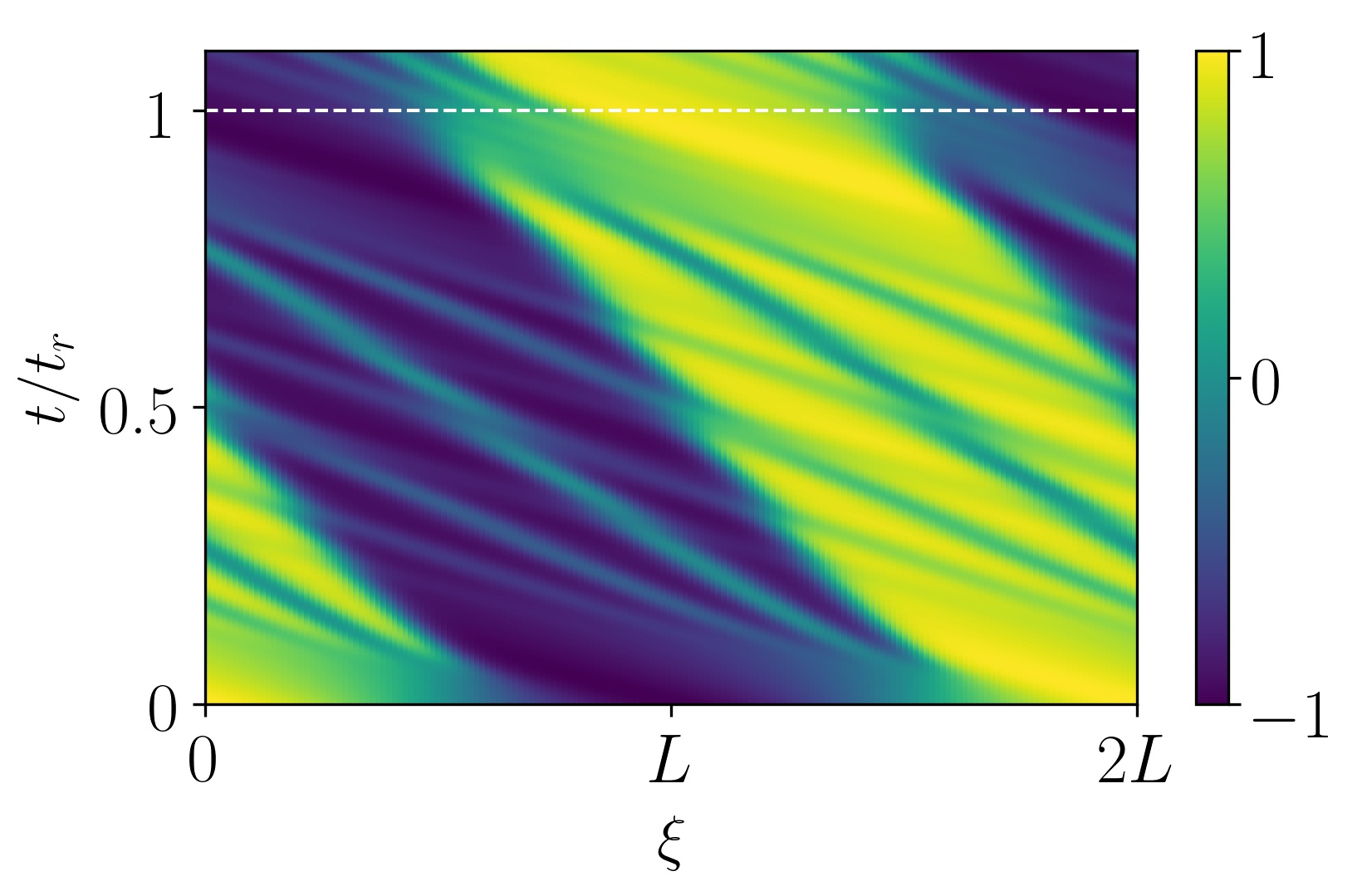}
	}
\caption{(color online) Shows a spacetime plot for the solution of the mKdV equation with an initial condition $\sqrt{\left|S\right|}L^{-1}\cos(\kappa \xi)$ such that (a) $S=50$, (b) $S=100$, (c) $S=-50$, and (d) $S=-100$. The time axis is normalized by the numerically measured FPUT recurrence time on the lattice. The color of the ``heat maps'' correspond to the value of $\phi L\left|S\right|^{-1/2}$ at that point in space and time. For both signs of beta, solitons form for these large values of $\left|S\right|$.}
\label{st_solitons}
\end{figure*}

\section{Role of mKdV solitons in the FPUT recurrence}\label{sec:mkdv}

Since Zabusky and Kruskal's semi-quantitative explanation of the $\alpha$-FPUT recurrence in terms of solitons in the Korteweg-de Vries (KdV) equation \cite{Zabusky1965}, it has been found that for $R\gtrsim 4$ in the $\alpha$-FPUT chain \cite{C.Y.Lin1999},  the FPUT lattice recurrence time agrees with the recurrence time in the continuum \cite{Zabusky1969}. Further, in the highly nonlinear regime solitons dominate the time evolution of the KdV equation \cite{Osborne1986}. As the value of $R$ increases, the number of solitons which form increases as well \cite{Salupere1996}. Before turning to an analytical estimate of the $\beta$-FPUT recurrence time in the highly nonlinear regime (large $\left|S\right|$) in section \ref{NLR}, we first investigate numerically the mKdV equation with initial conditions $A\kappa\cos(\kappa \xi) /2$ to see whether the mKdV solitons contribute similarly to the FPUT recurrence in the $\beta$-FPUT chain, as the KdV solitons do in the $\alpha$-FPUT chain. We consider large $N$ and set $\left|\beta\right|=1$, so the initial conditions can be written as $\sqrt{\left|S\right|}L^{-1}\cos(\kappa \xi)$. We note that the mKdV equation has both soliton and antisoliton solutions (see Appendix \ref{sec:mkdvsols}) while the KdV equation has only soliton solutions \cite{Zabusky1967} We numerically solve the mKdV equation by first replacing the spatial derivatives with finite differences that have an error of $\mathcal{O}(a^{4})$ \cite{Fornberg1988}. We then use the fourth-order Runge-Kutta method to find the time evolution. The lattice spacing is set to $a=1/(N+1)$, where $N=255$, and the time step size $dt=0.1$.

Figure \ref{st_solitons} shows the numerically determined solutions of the mKdV equation in a ``heat map'' with the horizontal axis being the spatial coordinate, the vertical axis being the time coordinate, and the color indicating the amplitude of the field. The field $\phi$ has been rescaled as $\phi L\left|S\right|^{-1/2}$ such that $1$ on the ``heat map'' corresponds to the maximum of the initial condition, and time, $t$, has been rescaled as $t/t_{r}$, where $t_{r}$ is the FPUT recurrence time found in section \ref{sec:numerics_beta_rec_scale} for the given $S$. A horizontal white dashed line is graphed at $t=t_{r}$ to emphasize the numerically determined FPUT recurrence time. We note that each value of $S$ in parts (a)-(d) of Figure \ref{st_solitons} is large enough that the FPUT recurrence time is proportional to $\left|S\right|^{-1/2}$. As in the previous section, we treat the $\beta>0$ and $\beta<0$ cases separately. 

Figures \ref{st_solitons}{\color{blue}{a}} and \ref{st_solitons}{\color{blue}{b}} show $S=50$ and $S=100$, respectively, with $\beta=1$ and $N=255$. The spacetime diagrams both show that shortly after $t=0$, solitons form on the bright background and antisolitons form on the dark background. Interestingly, for every soliton that forms there is an antisoliton that forms with the same velocity and absolute value of the amplitude. In other words, the solitons and antisolitons come in pairs. At $t=t_{r}$, all of the solitons (antisolitons) overlap on the bright (dark) background. These values of $\xi$ where the solitons (antisolitons) overlap correspond to roughly the same point where the (antisolitons) solitons initially formed on the (dark) bright background at $t=0$; therefore, there is a recurrence to the initial state. This is indeed similar to the mechanism which causes a recurrence of the initial state in the KdV equation \cite{Toda1975}. Increasing $S$ causes more solitons to form and does not affect the agreement of the recurrence time to that found on the lattice. However, for larger $S$ the increased number of solitons makes it less likely that they will all successfully overlap at the same point, thereby weakening the recurrence. This is evident when $S=100$ at $t=t_{r}$ where there is a single soliton and antisoliton not overlapping with their respective others.

Considering next $\beta<0$, Figures \ref{st_solitons}{\color{blue}{c}} and \ref{st_solitons}{\color{blue}{d}} show $S=-50$ and $S=-100$, respectively, where the only thing that has changed from the previous paragraph is the sign of $\beta$. Immediately after $t=0$, once again we observe that pairs of antisolitons and solitons form. However, unlike the $\beta>0$ case, the antisolitons form on the bright background and the solitons on the dark background. Also, whereas the (anti)solitons for $\beta>0$ traveled to the right, opposite to the background which traveled to the left, the (anti)solitons for $\beta<0$ travel to the left along with the background. A more significant difference which affects the recurrence is that the $\beta<0$ solitary waves are always soliton-like when on the dark background and antisoliton-like when on the bright background (which is why $-1\leq\phi L\left|S\right|^{-1/2}\leq 1$). For example, a soliton that is initially on the dark background will always be soliton-like while on the dark background, but if it travels to the bright background, it becomes antisoliton-like. This did not happen for $\beta>0$: Figures \ref{st_solitons}{\color{blue}{a}} and \ref{st_solitons}{\color{blue}{b}} show that the (anti)solitons were essentially unaffected by their background. At the FPUT recurrence time, as for $\beta>0$, all of the solitons overlap at one point and all of the antisolitons overlap at another point, which gives rise to approximately the initial conditions. However, the (anti)solitons that all meet at the same point at $t=t_{r}$ are not all the same (anti)solitons which formed together at the same point soon after $t=0$. This is because the solitary waves change their ``form'' (soliton to antisoliton or antisoliton to soliton) based on their background. This difference in how the solitons for $\beta<0$ interact with the background causes the recurrence to occur sooner. Indeed, for $\beta>0$, close inspection of Figures \ref{st_solitons}{\color{blue}{a}} and \ref{st_solitons}{\color{blue}{b}} shows that at $t\sim t_{r}/2$, there are two points in space, $\xi=0$ and $\xi=L$, where all of the solitons and antisolitons are overlapping. However, there are both solitons and antisolitons at these points, and therefore, a recurrence does not occur. This cannot happen in the $\beta<0$ case because there are never antisoliton-soliton interactions due to solitons (antisolitons) only ever existing on the dark (bright) background.

An important difference between positive and negative $\beta$ that in fact provides an explanation for our numerical results is that the negative $\beta$ mKdV has {\it kink} solutions, while the positive $\beta$ case does not. As shown in previous studies (see reference \cite{Perelman1974}) when a soliton interacts with a kink, the interaction causes the soliton to flip signs, thereby becoming an antisoliton. Therefore, given the numerical results, we argue that the background for the $\beta<0$ mKdV equation becomes a traveling kink-antikink. This is further supported by the form the background takes for negative $\beta$: nearly a constant value until it very quickly changes from bright to dark or dark to bright. In addition, this interpretation is also supported by the observation that negative $\beta$ solitons and the background both undergo a phase shift when the soliton changes its background, just as in soliton-kink interactions \cite{Gardner1995}.

\section{Analytics in the Highly Nonlinear Regime}\label{NLR}

In the highly nonlinear regime of the $\alpha$-FPUT chain, the FPUT recurrence time was analytically estimated from the velocities of the KdV solitons, both neglecting \cite{Toda1975} and including \cite{Wedding1982,Osborne1986} phase shifts due to soliton-soliton interactions. As seen from the numerics in the previous section, in the highly nonlinear regime of the $\beta$-FPUT chain, the recurrence of the initial state in the mKdV equation is also due to the solitons and antisolitons overlapping in their initial configuration. In this section, we seek an analytic estimate of the FPUT recurrence time by considering the soliton velocities. To do so, we first rewrite the mKdV equation in the ``Lax pair'' formalism \cite{Lax1968}
\begin{equation}
\mathcal{L}_{\tau} = \left[\mathcal{A},\mathcal{L}\right], 
\end{equation}
where $\left[\hspace{5pt},\hspace{5pt}\right]$ is the commutator and the Lax pair ($\mathcal{L}$ and $\mathcal{A}$) is given by \cite{Aktosun2011}
\begin{align}
\mathcal{L} &\equiv 
i \begin{pmatrix}
1 & 0 \\
0 & -1
\end{pmatrix}\partial_{\xi} - \frac{i\phi}{\sqrt{ 6b\zeta}}
\begin{pmatrix}
0 & 1 \\
1 & 0
\end{pmatrix},\\
\mathcal{A} &\equiv  
 -4\zeta
\begin{pmatrix}
1 & 0 \\
0 & 1
\end{pmatrix} \partial_{\xi}^{3} -b
\begin{pmatrix}
\phi^{2} & -\phi_{\xi}\sqrt{6b\zeta} \\
\phi_{\xi}\sqrt{6b\zeta} & \phi^{2}
\end{pmatrix} \partial_{\xi}\nonumber\\
&\hspace{40pt}-
\frac{b}{2}
\begin{pmatrix}
2\phi\phi_{\xi} & -\phi_{\xi\xi}\sqrt{6b\zeta} \\
\phi_{\xi\xi}\sqrt{ 6b\zeta} & 2\phi\phi_{\xi}
\end{pmatrix}  .
\end{align}
The eigenvalue problem $\mathcal{L}\vec{\psi}(\xi)=\sqrt{E} \vec{\psi}(\xi)$ is a $1+1$ dimensional Dirac equation
\begin{equation}\label{Dirac_mKdV}
\pm i \left(\psi_{\pm}\right)_{\xi}-\frac{i\phi}{\sqrt{6b\zeta}}\psi_{\mp} = \sqrt{E}\psi_{\pm}
\end{equation}
which has been used to show that the mKdV equation can be solved exactly by the inverse scattering transform \cite{Wadati1972}. Introducing the variables $\Psi_{\pm} = \psi_{-}\pm i\psi_{+}$, we can rewrite this Dirac equation as two Schr\"{o}dinger equations \cite{Lamb1980}
\begin{equation}\label{mkdvshro}
-(\Psi_{\pm})_{\xi\xi}+\left(-\frac{b}{6\zeta}\phi^{2} \pm i\frac{1}{\sqrt{6b\zeta}}\phi_{\xi}\right)\Psi_{\pm} = E \Psi_{\pm}.
\end{equation}
We note that equations (\ref{mkdvshro}) represent the corresponding Schr\"{o}dinger equations of the KdV equation written in terms of the solutions to the mKdV equation through the Miura transform \cite{Miura1968}. Notice, however, that the Hamiltonians in the Schr\"{o}dinger equations for $\beta >0$ are non-hermitian! However, if $\phi(\xi) = \phi(-\xi)$, then the potential is parity-time ($\mathcal{PT}$) symmetric and the spectra, $E$, can still be entirely real \cite{Bender2005, Bender2018, Wadati2008}.

Let us point out that $E$ is a \textit{time-independent} parameter and $\phi$ satisfies the mKdV equation at \textit{any} fixed time. Therefore, we can fix time to when the solitons have formed and then solve for $E$ in terms of their velocities. Then, we can go back to $\tau=0$ and relate the solitons' velocities to the parameters in the $\beta$-FPUT chain. If we neglect any phase shifts due to soliton-soliton, antisoliton-soliton, and antisoliton-antisoliton interactions, the FPUT recurrence occurs when the difference in speed between two consecutive (anti)solitons multiplied by the FPUT recurrence time is equal to $2L$ for $\beta>0$ and $L$ for $\beta<0$. The factor of two difference comes from the numerical results presented in section \ref{sec:mkdv}. For $\beta>0$, at the FPUT recurrence time, every initially formed soliton will be overlapping at about the same point, and therefore consecutive solitons, due to periodic boundary conditions, will be spaced out by the length of the space, $2L$. For $\beta<0$, close inspection of Figures \ref{st_solitons}{\color{blue}{c}} and \ref{st_solitons}{\color{blue}{d}} shows that at the FPUT recurrence time, the spacing between the first and second soliton will instead be the distance between the two overlapping points, which from the numerics is half of the space, $L$. This is a direct consequence of the background for negative $\beta$ becoming a kink-antikink, as discussed at the end of section \ref{sec:mkdv}. 

Denoting the difference in speed between the $(n+1)^{\text{st}}$ and the $n^{\text{th}}$ soliton by $\Delta v$, the FPUT recurrence time for the mKdV equation is therefore given by
\begin{equation}\label{approx_rec_time_idea}
\tau_{r} = \frac{\left(b+3\right)}{2}\frac{L}{\Delta v}.
\end{equation}
The numerical coefficient is chosen such that for $\beta>0$ ($\beta<0$), then $b=1$ ($b=-1$) and it takes the value $2$ ($1$).

Since we are neglecting phase shifts due to the soliton interactions, we only need to consider the one-soliton solution of the mKdV equation on the real line. Furthermore, we only need to keep track of the solitons; since each soliton is paired with an antisoliton, when the solitons are distributed in space such that the FPUT recurrence occurs, so will be the antisolitons. Appendix \ref{sec:mkdvsols} shows the explicit form of the soliton solutions for our normalization of the mKdV equation and also shows that for both positive and negative $\beta$, the difference in speed between consecutive solitons, $\Delta v\equiv |v_{n+1}-v_{n}|$, is
\begin{equation}\label{speed_diff}
\Delta v = 4\zeta\left|E_{n+1}-E_{n}\right|.
\end{equation}

Now considering $\tau=0$, using the initial conditions given in section \ref{chain_and_cont}, the corresponding stationary Schr\"{o}dinger equation is
\begin{equation}\label{mkdv_shro_0}
-(\Psi_{\pm})_{\xi\xi}\hspace{-2pt}+\hspace{-2pt}\left(-b\frac{\kappa^{2}A^{2}\cos^{2}(\kappa \xi) }{24\zeta}\hspace{-2pt}\mp\hspace{-2pt}\frac{i}{\sqrt{b}}\frac{\kappa^{2}A\sin(\kappa \xi)}{2\sqrt{6\zeta}}\right)\hspace{-2pt}\Psi_{\pm} \hspace{-2pt}=\hspace{-2pt} E \Psi_{\pm}.
\end{equation}
We proceed (following Toda \cite{Toda1975}) 
by making the harmonic approximation and Taylor expanding to second order in $\xi$. Focusing first on $\beta>0$, we expand about zero and, after some algebraic manipulations, find that the Schr\"{o}dinger equation becomes
\begin{equation}\label{mkdv_shro_1}
-\left(\Psi_{\pm}\right)_{\xi\xi}+\frac{A^{2}\kappa^{4}}{24\zeta}\hspace{-2pt}\left(\hspace{-2pt}\xi\hspace{-2pt} \mp\hspace{-2pt} i\frac{\sqrt{6\zeta}}{\kappa A}\right)^{2}\hspace{-2pt}\Psi_{\pm} \hspace{-2pt}=\hspace{-2pt} \left(\hspace{-2pt}E\hspace{-2pt}+\hspace{-2pt}\frac{\kappa^{2}A^{2}}{24\zeta}\hspace{-2pt}-\hspace{-2pt}\frac{\kappa^{2}}{4}\right)\hspace{-2pt}\Psi_{\pm},
\end{equation}
which is a shifted quantum harmonic oscillator. Despite the shift being an imaginary number, because the boundary conditions are the same as if the shift were a real number, the multiplicative constant on the right-hand side is equal to the spectra of the harmonic oscillator \cite{Znojil1999}
\begin{equation}\label{spect}
E_{n}+\frac{\kappa^{2}A^{2}}{24\zeta}-\frac{\kappa^{2}}{4} = \frac{A\kappa^{2}}{\sqrt{6\zeta}}\left(n+\frac{1}{2}\right),
\end{equation}
with the index $n$ labeling the solitons. Therefore, from equations (\ref{approx_rec_time_idea}) and (\ref{speed_diff}), the recurrence time is given by
\begin{equation}
\tau_{r} = \frac{L}{A\kappa^{2}}\sqrt{\frac{3}{2\zeta}}.
\end{equation}
Rewriting all of the mKdV parameters in terms of the $\beta$-FPUT chain's parameters, the rescaled recurrence time is found to be
\begin{equation}
T_{r} = \frac{2\sqrt{6}}{\pi^{2}\sqrt{\left|\beta\right|} A}.
\end{equation}
Using equation (\ref{Namp}) in the appendix, we can express the amplitude in terms of the parameter $S$ in the large $N$ limit and find
\begin{equation}\label{pos_beta_soliton_estimate}
T_{r} = \frac{\sqrt{6}}{\pi}\left|S\right|^{-1/2}.
\end{equation}

For the $\beta<0$ case, we expand about $\xi = \pm\pi/(2\kappa)$ and the corresponding Schr\"{o}dinger equation becomes
\begin{equation}\label{mkdv_shro_2}
-\left(\Psi_{\pm}\right)_{\xi\xi}+\frac{(A^{2}+A\sqrt{6\zeta})\kappa^{4}}{24\zeta}\hspace{-2pt}\left(\hspace{-2pt}\xi\hspace{-2pt} \mp\hspace{-2pt} \frac{\pi}{2\kappa}\right)^{2}\hspace{-2pt}\Psi_{\pm} \hspace{-2pt}=\hspace{-2pt} \left(\hspace{-2pt}E\hspace{-2pt}+\hspace{-2pt}\frac{\kappa^{2}A}{2\sqrt{6\zeta}}\right)\hspace{-2pt}\Psi_{\pm}.
\end{equation}
Therefore for $\beta<0$, the spectra follows
\begin{equation}
E_{n}+\frac{\kappa^{2}A}{2\sqrt{6\zeta}} = \kappa^{2}\sqrt{\frac{A^{2}+A\sqrt{6\zeta}}{6\zeta}}\left(n+\frac{1}{2}\right),
\end{equation}
and carrying through the calculation as done for $\beta>0$, we find
\begin{equation}\label{trecneg}
T_{r} = \frac{3\sqrt{2}}{\pi\sqrt{12|S|+\pi\sqrt{6|S|}}}.
\end{equation}

 \begin{figure}[t!]
		\centering
		\includegraphics[width=.48\textwidth]{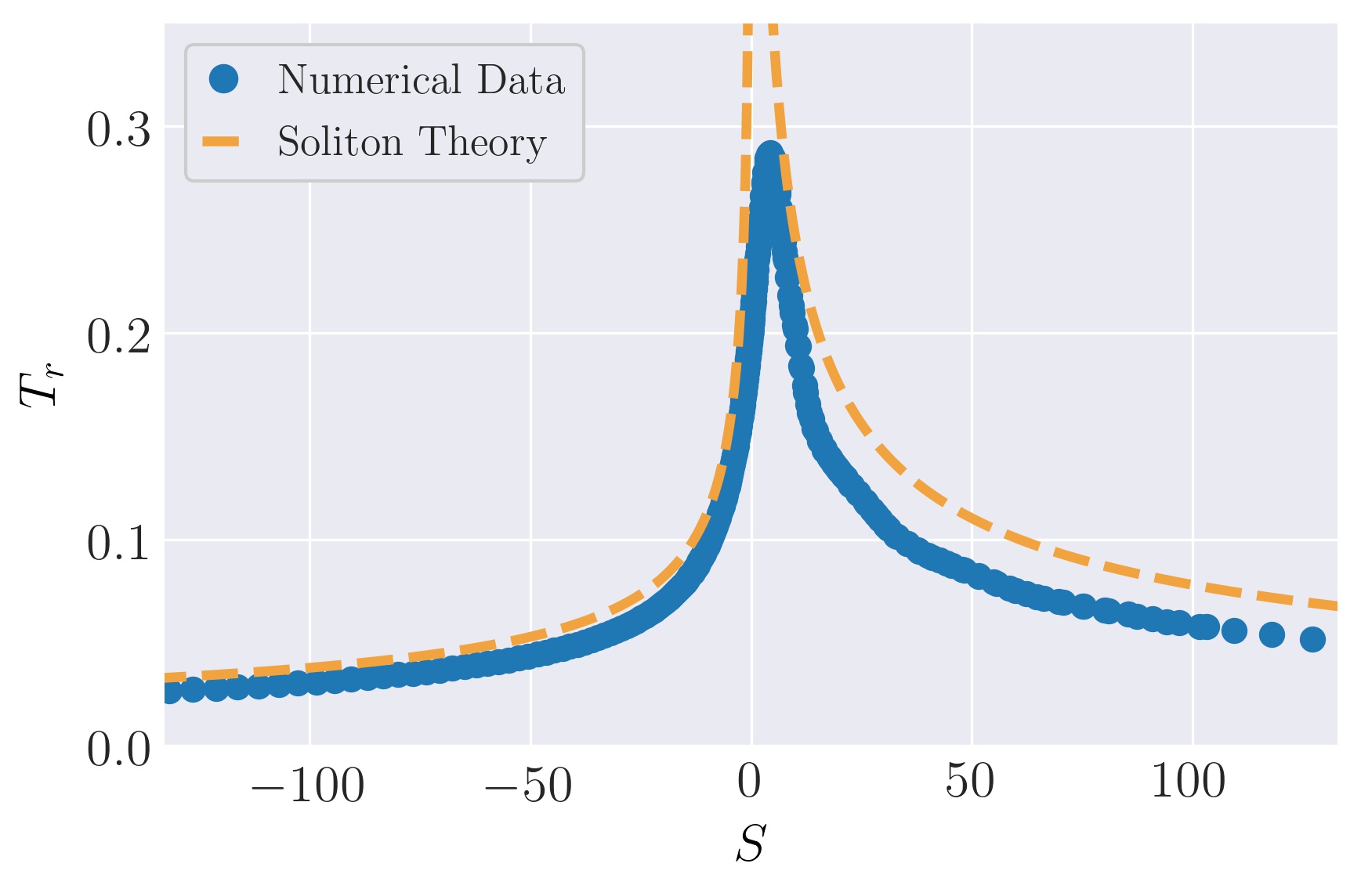}
		\caption{A comparison of the numerically generated rescaled recurrence time and analytical expressions obtained from the non-interacting soliton approximations which are given by equations (\ref{pos_beta_soliton_estimate}) and (\ref{trecneg}).} 
		\label{analytic_beta_rec_scale}
\end{figure}

Thus, we can see that for large $\left|S\right|$, the rescaled recurrence time goes like $|S|^{-1/2}$, as predicted by the numerics in section \ref{sec:numerics_beta_rec_scale}. The factor in front of $|S|^{-1/2}$ for $\beta>0$ ($\beta<0$) is about $1.33$ ($1.27$) times larger than the numerical factor in equation (\ref{eqn:pos_beta_scale}) (equation (\ref{eqn:neg_beta_scale})). This discrepancy is due to the harmonic approximation used to simplify equation (\ref{mkdv_shro_0}) along with neglecting phase shifts in our calculation. We note that Wedding and J\"{a}ger performed similar calculations with the KdV equation where they approximated $\cos(x)\sim 4\sech^{2}\left(x/\sqrt{8}\right)-3$ and included first-order corrections to the recurrence time due to phase shifts \cite{Wedding1982}. Such a calculation should improve the multiplicative constant but is outside the scope of the current work.

\begin{figure}[t!]
		\centering
		\includegraphics[width=.48\textwidth]{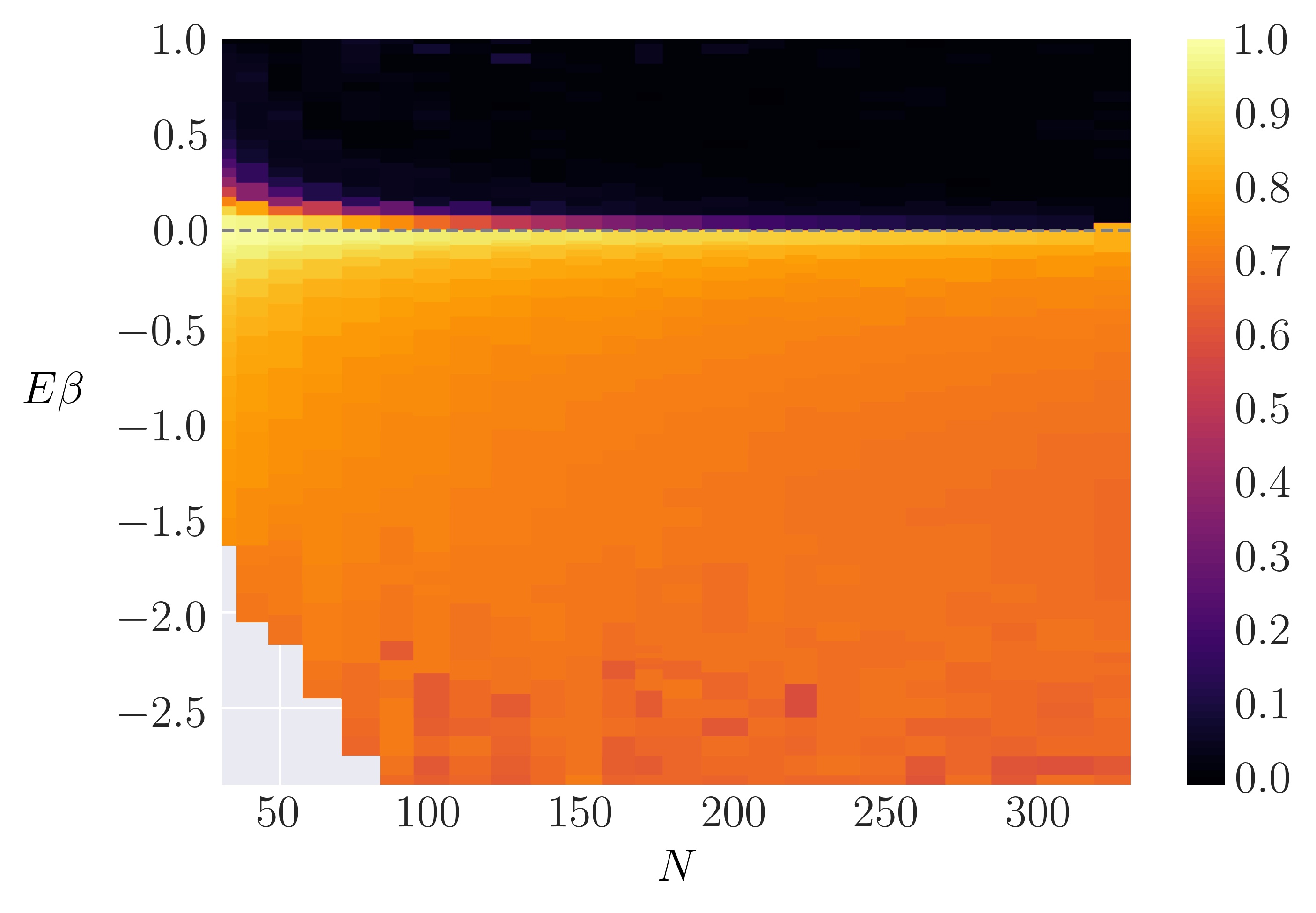}
		\caption{(color online) A plot of $E_1^{min}/E$, as a function of both $N$ and $E\beta$. This represents the energy that never leaves the first normal mode for $t < t_{max}$ and is therefore not considered when computing shareable energy. The gray area represents systems that blow up before $t_r$.} 
		\label{MinEnergy}
\end{figure}

\section{Existence of FPUT Recurrences} \label{sec:rec_existence}

After understanding the scaling of the FPUT recurrence time as functions of $E$, $\beta$, and $N$, we now explore which values of these parameters cause the FPUT recurrences to form. To compare the formation of an FPUT recurrence with an absence thereof, we investigate $E_{1}^{max}$, the maximum energy in the first normal mode on the interval $0.5t_r<t<1.5t_r$, with $t_r$ given by the numerical scalings established in section \ref{sec:numerics_beta_rec_scale}. We will denote the time at which $E_1^{max}$ occurs as $t_{max}$. When an FPUT recurrence forms, we expect $t_{max}$ to be the precise value of $t_r$.

To quantify the formation of an FPUT recurrence, we look at the proportion of ``shareable energy'' in the first normal mode at $t=t_{max}$. We define shareable energy as $\mathcal{E}(t)\equiv E_{1}(t)-E_1^{min}$, where $E_1^{min}$ is the minimum energy present in the first normal mode in the interval $0<t<t_{max}$. $\mathcal{E}$ is defined this way so it does not include the energy that is locked in the first normal mode. This is necessary to compare $\beta<0$ and $\beta>0$ because, as shown in Figure \ref{MinEnergy}, while most of the energy initially leaves the first normal mode at some time in the region $0<t<t_r$ for $\beta>0$ (as indicated by the black region), nearly $70\%$ of it stays in the first normal mode for $\beta<0$ (as indicated by the orange colored region). 

\begin{figure}[t!]
		\centering
		\includegraphics[width=.48\textwidth]{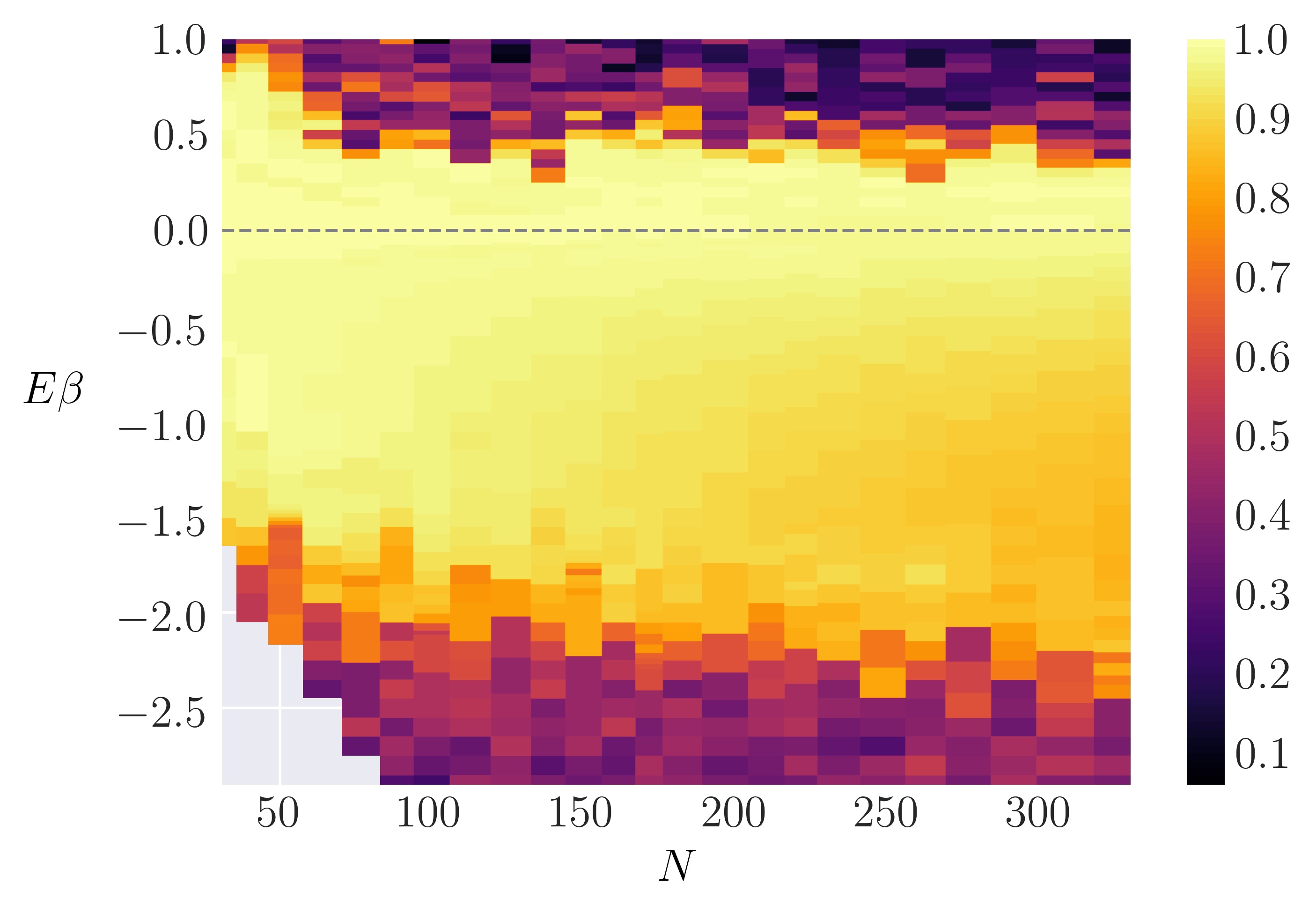}
		\caption{(color online) A ``heatmap" of the ``quality'' of recurrences in the $\beta$ FPUT model, for various system sizes $N$ and initial conditions of constant $E\beta$ in the first normal mode. The quantity plotted is $\mathcal{E}(t_{max})/\mathcal{E}(0)$. This quantity represents the proportion of ``shareable'' energy that returns to the initial conditions. The gray area represents systems that blow up before $t_r$.} 
		\label{heatmap}
\end{figure}

When an FPUT recurrence forms, we expect the proportion of shareable energy at $t_{max}$, $\mathcal{E}(t_{max})/\mathcal{E}(0)$, to be nearly $1$, whereas when no recurrence forms, the proportion of shareable energy should be much smaller than $1$. Figure \ref{heatmap} is a ``heat map'' of $\mathcal{E}(t_{max})/\mathcal{E}(0)$ for the same values of $E\beta$ and $N$ as Figure \ref{MinEnergy}. It is clear that for large $N$ there is a critical value of $E\beta$ where recurrences stop forming. Considering $63\leq N\leq330$, for $\beta>0$ this $E\beta_{c}^+ = 0.53\pm.035$ whereas for $\beta<0$, $E\beta_{c}^- = -2.43\pm.055$ (95\% confidence interval). Surprisingly, the magnitude of $E\beta_{c}^-$ is almost five times greater than the magnitude of $E\beta_{c}^+$, meaning FPUT recurrences exist for much higher initial energies in the negative $\beta$ case. It can be seen from equation (\ref{blowup}) in the appendix that for large $N$, the $\beta<0$ model can experience blow-up for initial $E\beta \lesssim -0.25$. It is interesting to note that for $\beta<0$, FPUT recurrences exist for initial energies up to nearly 10 times the energy above which blow-up is possible.

We can explain why the critical value at which FPUT recurrences stop forming is dependent on the energy but not the system size for large $N$ by considering the so-called ``metastable state,'' which is an out-of-equilibrium quasi-stationary state that prevents (for a given period of time) the system from thermalizing \cite{Benettin2008}. Indeed, previous numerical investigations by Livi and Ruffo \cite{Livi1992} for $\beta>0$ used the spectral entropy to show that at a fixed time and large $N$, the metastable state breaks down at an $N$-independent value $E\beta_{c}^+\sim \mathcal{O}\left(10^{-1}\right)$. Later, De Luca \textit{et al.} used their four-mode subsystem to  show analytically that for $S\gg 1$ in the $\beta>0$ FPUT chain, $E\beta_{c}^+=0.28$ is a critical energy for the diffusion of energy from low to high normal modes \cite{DeLuca1995}. There has yet to be a study, at least to our knowledge, on these subjects for $\beta<0$. Nevertheless, we would expect, given the numerics presented in Figure \ref{heatmap}, that the explanation would be similar. These results suggest that in the thermodynamic limit ($N\to\infty$), where $E\to\infty$ due to extensivity, FPUT recurrences will never form. We stress that this involves the breakdown of the metastable state, not the timescale to equipartition. Indeed, previous studies have found that the time it takes to achieve energy equipartition is dependent on the energy density, $E\beta/N$ \cite{Lvov2018}. The same Livi and Ruffo study \cite{Livi1992} indeed found that at a fixed time, energy equipartition is achieved at a critical value of $E\beta/N$.

To ensure that our results are not a numerical artifact, we have confirmed that the numerics preserve time-reversal symmetry at, and slightly above, the critical value of $E\beta$. For $\beta<0$, our results were time-reversible for all $E\beta$ and $N$, with a time step size $dt = 0.1$. However, for $\beta >0$, for large $N$ and $E\beta \gg E\beta_{c}^+$, we were not able to ensure time reversibility with our allotted computational resources. This is due to the exponential instabilities which have been noted in other studies for $\beta>0$ \cite{Drago1987, Pace2019} and has been explicitly studied as well in \cite{Driscoll1976}. We have confirmed that lowering the time step size, even if it is not small enough to ensure time-reversibility, does not change the qualitative properties of Figure \ref{heatmap}.

\section{Conclusions and Discussion}\label{sec:conclusion}

Despite the comprehensive studies FPUT recurrences in the $\alpha$-FPUT chain have received in the past, there has not previously been a similarly comprehensive study of the FPUT recurrences in the $\beta$-FPUT chain. Using both numerical and analytical methods, we have examined these FPUT recurrences for \textit{both} $\beta<0$ and $\beta>0$ in the $\beta$-FPUT chain and its continuum limit, the modified Korteweg-de Vries (mKdV) equation. Our results establish that the rescaled FPUT recurrence time $T_{r} = t_{r}/(N+1)^{3}$ is, for large $N$, completely described by the parameter $S\equiv E\beta(N+1)$, and we have investigated the dependence of the FPUT recurrences on this and other lattice parameters.

For small $\left|S\right|$, the lattice is in a nearly linear regime. By numerically integrating the lattice model, $T_{r}$ was found to transition smoothly between $\beta<0$ and $\beta>0$ as a function of $S$. Interestingly, the maximum value of $T_{r}$ is located at $S\sim 4.2$. Using the shifted-frequency perturbation theory, we extended results from reference \cite{Sholl1991} and found a closed form for $T_{r}$, given in equation  (\ref{pert_rec_time}), which becomes a function of only $S$ with an error of order $\mathcal{O}\left((N+1)^{-2}\right)$. This expression was found to accurately describe $T_{r}$ for $-12\lesssim S\lesssim 4$.

In the highly nonlinear regime (large $\left|S\right|$), our numerical investigations revealed that for both $\beta<0$ and $\beta>0$, $T_{r}\propto \left|S\right|^{-1/2}$. However, the numerically fitted constant for $\beta>0$ was $0.5862$ while it was $0.3078$ for $\beta<0$, and $T_{r}$ followed this power-law when $S\gtrsim 30$ and $S\lesssim -12$. We then went to the continuum limit and numerically investigated the role of mKdV antisolitons and solitons in the recurrence to the initial state. We found that as in the continuum limit of the $\alpha$-FPUT chain, solitons form and strongly influence the temporal evolution in the highly nonlinear regime. In this soliton-dominated regime, we found that $T_{r}\propto \left|S\right|^{-1/2}$ agrees well with the recurrence time in the continuum. Furthermore, the recurrence was found to occur sooner for negative $\beta$ due to soliton-kink interactions which do not occur for positive $\beta$. Finally, in the highly nonlinear regime, we estimated the FPUT recurrence time analytically by considering the velocities of the solitons and correctly replicated the $\left|S\right|^{-1/2}$ power-law scaling of the rescaled FPUT recurrence time for both $\beta<0$ and $\beta>0$.

Lastly, we concluded our study with investigations into which lattice parameters cause FPUT recurrences to form. We found that for large enough $N$, there is a critical value of $E\beta$ above which FPUT recurrences do not form. The energies are, for $\beta<0$ and $\beta>0$ respectively, $E\beta_{c}^- = -2.43 \pm .055$ and $E\beta_{c}^+ = 0.53 \pm .035$, which are surprisingly quite different. This critical value of $E\beta$ has been associated with the breakdown/lack of formation of the metastable state. Another interesting result of this section was the discovery that while nearly all of the energy leaves the first normal mode before $t_{max}$ for $\beta>0$, roughly $70\%$ of the energy remains ``locked'' in the first normal mode up to $t_{max}$ for $\beta<0$. 

While there have been numerous studies on the stability and localizing properties of $q$-breathers \cite{Flach2005,Flach2006, Penati2007,Flach2008} and $q$-tori \cite{Christodoulidi2010,Christodoulidi2013}, there has yet to be a quantitative comparison of their periods to the FPUT recurrence time. The results presented here for the $\beta$-FPUT chain and those from reference \cite{Lin1997} for the $\alpha$-FPUT chain provide an opportunity for a detailed study of this important open question. 

The results of the current study also strongly suggest that the $\beta$-FPUT chain needs to be further investigated for $\beta<0$ with energy small enough to avoid blow-up. Just as the FPUT recurrences behaved and scaled differently based on the sign of $\beta$, we expect other features of the dynamics to change with the sign of $\beta$. This includes the timescale to energy equipartition, which has been extensively studied in the $\beta>0$ case \cite{DeLuca1995,Livi1985,Berchialla2004,Benettin2011,Lvov2018}, and also the intermittent dynamics at equilibrium. Recent studies revealed the equilibrium dynamics in the $\alpha$-FPUT chain has long excursions from the equilibrium manifold due to sticky regions of phase space caused by $q$-breathers \cite{Danieli2017}.

As a final remark, we recall the comment in our introduction that the dynamics of a one-dimensional Bose gas in the quantum rotor regime maps onto the $\beta$-FPUT chain with $\beta<0$ \cite{Danshita2014}. This suggests that our results may be of interest in studies of ultra-cold bosons confined in optical lattices. 

\section{Acknowledgments}

We thank Felix Izrailev for stimulating discussions and Mark Ablowitz for valuable insights into the solutions of mKdV equation. S.D.P. and K.A.R. are grateful for support from Boston University's Undergraduate Research Opportunities Program. Finally, we thank Boston University's Research Computing Services for their computational resources.
\begin{appendix}

\section{Relationship Between Energy and Coordinate Space Initial Amplitudes} \label{amps}

The initial condition in real space considered in this study is $q_{n} = A\sin(n\pi/(N+1))$. From the canonical transformation, equation (\ref{canonical_transformation}), the initial normal mode coordinate is therefore $Q_{k}(0) = A\delta_{1,k}\sqrt{(N+1)/2}$. Using this initial condition and considering the Hamiltonian in normal mode coordinates, equation (\ref{beta_NM_hamiltonian}), the total energy of the $\beta$-FPUT chain is
\begin{equation}
\label{benergy}
E = \frac{\omega_{1}^{2} A^{2}(N+1)}{4} \left(1 + \frac{3b}{8}\omega_{1}^{2}A^{2}\left|\beta\right|\right).
\end{equation} 
Noting that $S\equiv E\beta(N+1)$ is the relevant parameter in the scaling of the $\beta$-FPUT recurrences, we rewrite equation (\ref{benergy}) as
\begin{equation}\label{beta_initial_S}
\left|S\right| = \frac{ A^{2}\omega_{1}^{2}\left|\beta\right|(N+1)^{2}}{4} \left(1+\frac{3b\left|\beta\right|\omega_{1}^{2}A^{2}}{8}\right).
\end{equation} 
Solving for $A$, and taking only the positive amplitude, we find
\begin{equation}
A = \csc\left(\frac{\pi}{2(N+1)}\right)\sqrt{\frac{-1+\sqrt{1+6b\left|S\right|(N+1)^{-2}}}{3b\left|\beta\right|}}.
\end{equation}
As we are interested in the limit of large $N$, we note that
\begin{equation}\label{Namp}
A = \frac{2}{\pi}\sqrt{\frac{\left|S\right|}{\left|\beta\right|}}+\mathcal{O}\left(\frac{1}{(N+1)^{2}}\right)
\end{equation}

\section{Small to Large System Size Threshold for $\bm{\beta<0}$ }\label{sec:beta_large_vs_small}
\begin{figure}[t!]
		\centering
		\includegraphics[width=.48\textwidth]{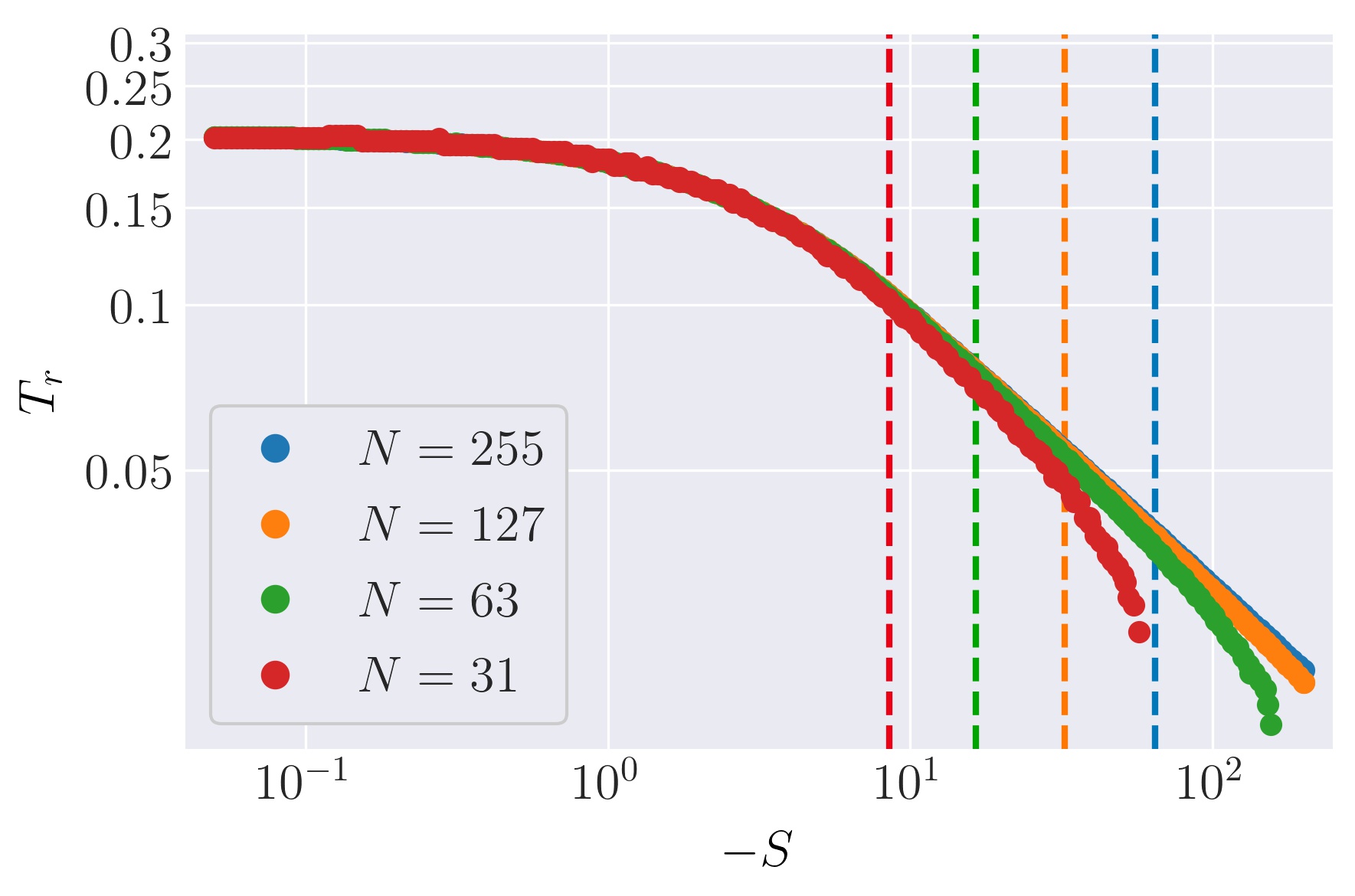}
		\caption{(color online) Shows the rescaled recurrence time as a function of $S$, but, unlike figure \ref{neg_beta_rec_scale}, also includes including data points when the system is too small and does not follow the large $N$ trend. The vertical lines are the critical $S$ between a small and large system which is given by equation (\ref{crit_S}).} 
		\label{small_size_trec}
\end{figure}

It has been shown that in the $\alpha$-FPUT chain the threshold for a system to be large enough to be described by its continuum limit is correlated to instabilities due to the potential being unbounded \cite{C.Y.Lin1999}. When $\beta<0$, the system is unbounded and thus we expect a similar threshold to exist. Letting $\Delta_{n}\equiv q_{n}-q_{n-1}$ and setting $\ddot{q}=\dot{q}=0$ in equation (\ref{beom}) with $\beta<0$, we find that the extrema satisfy
\begin{equation}
\Delta_{n}\left(1-|\beta|\Delta_{n}^{2}\right)=\Delta_{n+1}\left(1-|\beta|\Delta_{n+1}^{2}\right),
\end{equation}
which is essentially a cubic equation for $\Delta_{n+1}$ in terms of $\Delta_{n}$.
Letting $\Delta_{n}\equiv \Xi$ and solving for $\Delta_{n+1}$ , we find the three roots to be
\begin{align}
\Delta_{n+1} &=\Xi,\\
\Delta_{n+1} &\equiv \Delta^{(+)}= \frac{-\Xi|\beta|+\sqrt{|\beta|\left(4-3\Xi^{2}|\beta|\right)}}{2|\beta|},\\
\Delta_{n+1} &\equiv \Delta^{(-)}= \frac{-\Xi|\beta|-\sqrt{|\beta|\left(4-3\Xi^{2}|\beta|\right)}}{2|\beta|}.
\end{align}
From the condition that
\begin{equation}
\label{condition}
\sum_{n=1}^{N+1}\Delta_{n} = 0
\end{equation}
and letting there be $j_{1}$ springs with $\Delta^{(+)}$, $j_{2}$ springs with $\Delta^{(-)}$ and $N+1-j_{1}-j_{2}$ with $\Xi$, equation (\ref{condition}) becomes
\begin{equation}
j_{1}\Delta^{(+)}+j_{2}\Delta^{(-)}+\left(N+1-j_{1}-j_{2}\right)\Xi=0.
\end{equation}
Solving for $\Xi$, we find
\begin{equation}
\Xi=\frac{\pm(j_{1}-j_{2})|\beta|^{-1/2}}{\hspace{-8pt}\sqrt{3 (j_{1}^2\hspace{-1pt}+\hspace{-1pt}j_{2}^{2})\hspace{-2pt}+\hspace{-2pt}3 j_{1}j_{2}\hspace{-1pt}-\hspace{-1pt}3(j_{1}\hspace{-2pt}+\hspace{-1pt}j_{2}) (N\hspace{-2pt}+\hspace{-2pt}1)\hspace{-2pt}+\hspace{-1pt}(N\hspace{-2pt}+\hspace{-2pt}1)^2}}.
\end{equation}
We want to chose $j_{1}$, $j_{2}$, and the $\pm$ sign such that the energy
\begin{equation}\label{energy1}
\begin{aligned}
E_{c} &=j_{1}\left(\frac{1}{2}\left(\Delta_{n}^{(+)}\right)^{2}-\frac{|\beta|}{4}\left(\Delta_{n}^{(+)}\right)^{4}\right)\\
&\hspace{15pt}+j_{2}\left(\frac{1}{2}\left(\Delta_{n}^{(-)}\right)^{2}-\frac{|\beta|}{4}\left(\Delta_{n}^{(-)}\right)^{4}\right)\\
&\hspace{15pt}+\left(N+1-j_{1}-j_{2}\right)\left(\frac{1}{2}\Xi^{2}-\frac{|\beta|}{4}\Xi^{4}\right)
\end{aligned}
\end{equation}
is minimized, which will give us a lower estimate. The expression above is invariant under any of the $\Delta_{n}^{(+)}$, $\Delta_{n}^{(-)}$, or $\Xi$ variables picking up an overall minus sign. We also note that the expression for $\Xi$ picks up an overall minus sign when $j_{1}$ and $j_{2}$ are switched. Therefore, switching $j_{1}$ with $j_{2}$ in equation (\ref{energy1}) also causes $\Delta_{n}^{(-)}$ and $\Delta_{n}^{(+)}$ to switch, leaving the form for $E_{c}$ unchanged. Therefore, without loss of generality, we can set $j_{2}=0$. With this, the energy is minimized if we choose the $\pm$ sign in $\Xi$ to be $+$ and also chose $j_{1}=1$. The minimized energy threshold becomes
\begin{equation}
E_{c}=\frac{N^4-N+1}{4|\beta| \left(N^2-N+1\right)^2}.
\end{equation}
In terms of $S_{c}\equiv E_{c}\beta(N+1)$, we have
\begin{equation}\label{crit_S}
\left|S_{c}\right| = \frac{(N+1)(N^4-N+1)}{4\left(N^2-N+1\right)^2}.
\end{equation}
For large $N$, this goes like
\begin{equation}\label{blowup}
\left|S_{c}\right| \sim \frac{N+3}{4}+\mathcal{O}\left(\frac{1}{N}\right)
\end{equation}
Hence for $N\gtrsim 4\left|S\right|-3$, we expect the lattice to be considered large and thus to behave similarly to its continuum limit---the mKdV equation. Figure \ref{small_size_trec} shows the dependence of the rescaled FPUT recurrence times on $S$, as presented in section \ref{sec:numerics_beta_rec_scale} but also includes the numerical data that do not agree with the common trend shown in Figure \ref{neg_beta_rec_scale}. As discussed in the main text, this disagreement arises because the system sizes are too small to be described by the continuum limit.  The vertical lines represent the critical values of $S$ that separate the ``small" and  ``large" systems, as  given by equation (\ref{crit_S}). These lines are plotted for the same system sizes $N=31$, $63$, $127$, and $255$ and follow the same color coding as the numerical data. The numerical data starts to ``peel-off'' from its larger counterparts at values which are close to the vertical lines. Close inspection of the numerical data shows that disagreements with the larger counterparts occurs for $N=31$ at $S_{c}\sim7.1$, $N=63$ at $S_{c}\sim14.6$, and $N=127$ at $S_{c}\sim29.4$. This confirms that, as in the $\alpha$-FPUT chain \cite{C.Y.Lin1999}, there is indeed a correlation between the instabilities that arise due to saddle points and the small to large system size threshold.  

\section{Shifted-Frequency Perturbation Theory}\label{Sec:PT}
The underlying assumption of the shifted-frequency perturbation theory presented by Sholl and Henry in reference \cite{Sholl1991} is that that each normal mode coordinate can be written as
\begin{equation}\label{HS_PT}
Q_{k} =  \sum_{j=0}^{\infty}\beta^{j}Q_{k,j},
\end{equation}
and that there are nonlinear frequencies, defined by
\begin{equation}
\Omega_{k}^{2} \equiv \omega_{k}^{2} +\sum_{j=1}^{\infty} \beta^{j}\mu_{k,j}.
\end{equation}
The variables $Q_{k,j}$ are found by inserting equation (\ref{HS_PT}) into the normal mode coordinate equations of motion 
\begin{equation}
\ddot{Q}_{k}+\Omega_{k}^{2} Q_{k}=-\sum_{i,j,l=1}^{N} C_{kijl} Q_{i} Q_{j} Q_{l},
\end{equation}
where the $\omega_{k}^{2}$ in front of the $Q_{k}$ term on the left-hand side has been replaced by $\Omega_{k}^{2}$, but the coupling constant $C_{kijl}$ is still given in terms of the linear frequencies, as given in equation (\ref{coupling_constant}). After this substitution, the $\mu_{k,j}$s, called the ``frequency corrections,'' are then chosen so as to eliminate secular terms.

Sholl and Henry calculated up to third order corrections to $Q_{k}$ for initial conditions $Q_{k}(0)= \delta_{k,1}$, $P_{k}(0)=0$. We discussed the expression they found for the FPUT recurrence time in the $\beta$-FPUT chain in section \ref{sec:nearly_linear}. Here, we restate their results for the frequency corrections of the first, third, and fifth nonlinear frequency:
\begin{align}
\mu_{1,1} &= \frac{3}{4}C_{1111},\\
\mu_{1,2} &= -\frac{3}{4}C_{1111}A_{1,1}+\frac{3}{4}C_{1,1,1,3}\left(3A_{3,2}+A_{3,3}\right),\\
\mu_{3,1} &= \frac{3}{2}C_{3311},\\
\mu_{3,2}&=\frac{3}{4 A_{3,1}}\left[C_{3111}\left(2 B_{1,4}+B_{1,5}+B_{1,6}\right)\right. \nonumber\\ &\hspace{40pt}+C_{3311}\left(B_{3,5}+B_{3,6}-4 A_{3,1} A_{1,1}\right) \\ 
&\hspace{40pt}\left.+C_{3115}\left(2 B_{5,5}+B_{5,6}+B_{5,7}\right) \right],\nonumber \\
\mu_{5,1} &= \frac{3}{2}C_{5511}.
\end{align}
The $C_{kijl}$ are the coupling constants defined by equation (\ref{coupling_constant}), and the constants $A_{i,j}$ and $B_{i,j}$ are given by:
\begin{align}
\label{PT1}
A_{1,1}&= \frac{C_{1111}}{32\Omega_{1}^{2}},\\
A_{3,2}&= \frac{-3C_{1111}}{4\left(\Omega_{3}^{2}-\Omega_{1}^{2}\right)},\\
A_{3,3}&= \frac{-C_{1111}}{4\left(\Omega_{3}^{2}-9\Omega_{1}^{2}\right)},\\
A_{3,1}&= -A_{3,2}-A_{3,3},\\
B_{1,4}&=\frac{-3 C_{1113} A_{3}}{2\left(\Omega_{1}^{2}-\Omega_{3}^{2}\right)},\\
B_{1,5}&=\frac{-3 C_{1113} A_{3,1}}{4\left(\Omega_{1}^{2}-\left(\Omega_{3}-2 \Omega_{1}\right)^{2}\right)},\\
B_{1,6}&=\frac{-3 C_{1113} A_{3,1}}{4\left(\Omega_{1}^{2}-\left(\Omega_{3}+2 \Omega_{1}\right)^{2}\right)},\\
B_{3,5}&=\frac{-3 C_{3311} A_{3,1}}{4\left(\Omega_{3}^{2}-\left(\Omega_{3}-2 \Omega_{1}\right)^{2}\right)},\\
B_{3,6}&=\frac{-3 C_{3311} A_{3,1}}{\left(\Omega_{3}^{2}-\left(\Omega_{3}+2 \Omega_{1}\right)^{2}\right)}, \\
B_{5,5}&= \frac{3C_{5311}A_{3,1}}{2\left(\Omega_{3}^{2}-\Omega_{5}^{2}\right)},\\
B_{5,6}&= \frac{3C_{5311}A_{3,1}}{4\left(\left(\Omega_{3}-2\Omega_{1}\right)^{2}-\Omega_{5}^{2}\right)},\\
B_{5,7}&= \frac{3C_{5311}A_{3,1}}{4\left(\left(\Omega_{3}+2\Omega_{1}\right)^{2}-\Omega_{5}^{2}\right)}.
\end{align}
We note that constants $B_{5,5}$, $B_{5,6},$ and $B_{5,7}$ were never explicitly reported in reference \cite{Sholl1991}. They are found by solving equation (22) in their manuscript, which is a differential equation whose solution is the second-order corrections of the 5th normal mode coordinate, $Q_{5,2}$. Doing so, we find that
\begin{align}
Q_{5,2}(t) &= B_{5,1}\cos(\Omega_{5}t)+B_{5,2}\cos(\Omega_{1}t)+B_{5,3}\cos(3\Omega_{1}t)\nonumber\\
&+B_{5,4}\cos(5\Omega_{1}t)+B_{5,5}\cos(\Omega_{3}t)\nonumber\\
&+B_{5,6}\cos((2\Omega_{1}-\Omega_{3})t)+B_{5,7}\cos((2\Omega_{1}+\Omega_{3})t),
\end{align}
where $B_{5,5}$, $B_{5,6},$ and $B_{5,7}$ are listed above, and $B_{5,1}$, $B_{5,2}$, $B_{5,3}$, and $B_{5,4}$ are
\begin{align}
B_{5,1}&= -\sum_{n=2}^{7}B_{5,n},\\
B_{5,2}&= \frac{3(3A_{3,2}+A_{3,3})C_{5311}}{4(\Omega_{1}^{2}-\Omega_{5}^{2})},\\
B_{5,3}&= \frac{3(A_{3,2}+2A_{3,3})C_{5311}}{4(9\Omega_{1}^{2}-\Omega_{5}^{2})},\\
\label{PT2}
B_{5,4}&= \frac{3A_{3,3}C_{5311}}{4(25\Omega_{1}^{2}-\Omega_{5}^{2})}.
\end{align}
We note that in equations (\ref{PT1}-\ref{PT2}), all of the nonlinear frequencies are calculated to first order in $\beta$.

\section{Solitons in the mKdV equation}\label{sec:mkdvsols}

In section \ref{NLR}, we use the speed of the solitons which form in the continuum to find an approximate expression for the FPUT recurrence time on the lattice. In this Appendix, we restate known information in the literature about the one-soliton solutions of the mKdV equation given by equation (\ref{mkdvfinal}). For simplicity, we will not include the arbitrary phase shifts $\xi\to\xi+\xi_{0}$.

The one soliton solution on the real line is given by \cite{Perelman1974,Chanteur1987}
\begin{equation}\label{betasoliton}
\phi =\frac{3\sqrt{2}\left(\phi_{\infty}^{2}-v\right)}{\sqrt{3v-\phi^{2}_{\infty}}\cosh\left(\eta\sqrt{b(v-\phi_{\infty}^{2})}\right)-\phi_{\infty}\sqrt{2}}+\phi_{\infty},
\end{equation}
where $\eta = \zeta^{-1/2}(\xi-bv\tau)$ and $b=\text{sgn}(\beta)$. For $\beta<0$, solitons (antisolitons) reside on a negative (positive) background, while for $\beta>0$, solitons and antisolitons can be on both a positive or negative background. There are obvious restrictions on what value $\phi_{\infty}$ can take given the soliton speed. For $\beta>0$, the soliton exists for $-\sqrt{v}<\phi_{\infty}<\sqrt{v}$, while for $\beta<0$ it exists for $-\sqrt{3v}<\phi_{\infty}<-\sqrt{v}$. From this, it is clear that no one-soliton solutions exist at zero background for negative $\beta$. However, for positive $\beta$, we can set the background to zero and return to the well known result \cite{Zabusky1967, Wadati1972}
\begin{equation}
\label{mkdv_soliton}
\phi = \pm\sqrt{6v} \sech\left(\sqrt{\frac{v}{\zeta}}\left(\xi-v\tau\right)\right).
\end{equation}
Also, while it is not used in our results, we note that for $\beta>0$ there are also algebraic solitons \cite{Ono1976}
\begin{equation}
\phi = \sqrt{v}\left(1-\frac{12\zeta}{3\zeta+2v(\xi-v\tau)^{2}}\right).
\end{equation}

While there are solitons in both the $\beta>0$ and $\beta<0$ mKdV equations, the $\beta<0$ case also has kinks. The one-kink solution is \cite{Grosse1984}
\begin{equation}
\phi = \sqrt{3v}\tanh\left(\sqrt{\frac{v}{\zeta}}\left(\xi+v\tau\right)\right).
\end{equation}

The speed, $v$, of the soliton given in equation (\ref{betasoliton}) has been expressed in terms of the eigenvalue $E$ in the corresponding Schr\"{o}dinger equation given by equation (\ref{mkdvshro}) through the $N$-soliton solutions as \cite{Perelman1974,Chanteur1987}
\begin{equation}
v = \frac{\phi_{\infty}^{2}}{3}-4b\zeta E.
\end{equation}
Important for the calculation presented in section \ref{NLR}, the difference in the speed of consecutive solitons for both positive and negative $\beta$ is therefore given by
\begin{equation}
\Delta v = 4\zeta\left|E_{n+1}-E_{n}\right|. 
\end{equation}

\end{appendix}

\bibliographystyle{ieeetr}
\bibliography{B1oR_Notes}

\end{document}